\documentclass[12pt]{article} 

\RequirePackage[a4paper, margin=1in]{geometry}
\RequirePackage{amsthm,amsmath,amsfonts,amssymb}
\RequirePackage[authoryear]{natbib}
\RequirePackage{lipsum}
\RequirePackage[colorlinks,citecolor=blue,urlcolor=blue]{hyperref}
\RequirePackage{graphicx}
\RequirePackage[dvipsnames]{xcolor}
\RequirePackage{bm}
\RequirePackage{bbm}
\RequirePackage{setspace}
\RequirePackage[utf8]{inputenc}
\RequirePackage{parskip}
\RequirePackage{tabularray} 
\RequirePackage{titlesec}
\RequirePackage{caption}
\RequirePackage{subcaption}
\RequirePackage{placeins}
\RequirePackage{algorithm}
\RequirePackage{algorithmic}
\RequirePackage[T1]{fontenc}
\RequirePackage[affil-it]{authblk}

\DeclareUnicodeCharacter{03B3}{$\gamma$}

\DeclareMathOperator*{\argmin}{arg\,min}
\DeclareMathOperator*{\arginf}{arg\,inf}

\begin{document}
\onehalfspacing

\newcommand{\todo}[1]{\textcolor{red}{TODO: #1}}
\newcommand{\todomaybe}[1]{\textcolor{orange}{TODO?: #1}}
\newcommand{\note}[1]{\textcolor{magenta}{\textbf{NOTE}: #1}}
\newcommand{\aside}[1]{\textcolor{Plum}{\textbf{ASIDE}: #1}}
\newcommand{\question}[1]{\textcolor{OliveGreen}{\textbf{Question}: #1}}
\newcommand{\R}{\mathbb{R}}
\newcommand{\iid}{\overset{\text{iid}}{\sim}}
\renewcommand{\algorithmicrequire}{\textbf{Input:}}
\renewcommand{\algorithmicensure}{\textbf{Output:}}

\titleformat{\subsubsection}[runin] 
  {\normalfont\itshape} 
  {\thesubsubsection} 
  {1em} 
  {} 

\title{Simulation-based Bayesian inference under model misspecification}

\author[1,2]{Ryan P. Kelly}
\author[1,2]{David J. Warne}
\author[3]{David T. Frazier}
\author[4]{David J. Nott}
\author[5]{Michael U. Gutmann}
\author[1,2,6]{Christopher Drovandi}

\affil[1]{School of Mathematical Sciences, Queensland University of Technology, Australia}
\affil[2]{Centre for Data Science, Queensland University of Technology, Australia}
\affil[3]{Department of Econometrics and Business Statistics, Monash University, Australia}
\affil[4]{Department of Statistics and Data Science, National University of Singapore, Singapore}
\affil[5]{School of Informatics, Institute for Adaptive \& Neural Computation, University of Edinburgh, UK}
\affil[6]{Centre of Excellence for the Mathematical Analysis of Cellular Systems, Queensland University of Technology, Australia}


\maketitle

\begin{abstract}
Simulation-based Bayesian inference (SBI) methods are widely used for parameter estimation in complex models where evaluating the likelihood is challenging but generating simulations is relatively straightforward. However, these methods commonly assume that the simulation model accurately reflects the true data-generating process, an assumption that is frequently violated in realistic scenarios. In this paper, we focus on the challenges faced by SBI methods under model misspecification. We consolidate recent research aimed at mitigating the effects of misspecification, highlighting three key strategies: i) robust summary statistics, ii) generalised Bayesian inference, and iii) error modelling and adjustment parameters. To illustrate both the vulnerabilities of popular SBI methods and the effectiveness of misspecification-robust alternatives, we present empirical results on an illustrative example.

\end{abstract}


\section{Introduction}\label{sec:intro}



Standard Bayesian methods rely on explicitly defined likelihood functions derived from parametric statistical models. However, in many real-world applications, directly evaluating this likelihood can be computationally prohibitive or analytically intractable. In such scenarios, an \textit{implicit} statistical model \citep{diggle_monte_1984} can be used by specifying the data-generating process (DGP) directly.

Simulation-based Bayesian inference (SBI) methods allow for approximating the posterior distribution through simulations of implicit statistical models. Approximate Bayesian computation (ABC) methods generate simulated datasets under candidate parameter values and then computing a discrepancy—often defined through lower-dimensional summary statistics—between observed and simulated data \citep{tavare_inferring_1997, martin_approximating_2024, sisson_handbook_2018}.
Another long-established method is indirect inference, which estimates the parameters of a statistical model based on indirect or auxiliary summaries of the observed data \citep{gourieroux_indirect_1993}. A related and popular SBI technique is Bayesian synthetic likelihood (BSL) \citep{price_bayesian_2018, wood_statistical_2010}, which builds on indirect inference by assuming that these summary statistics follow a (conditionally) multivariate normal distribution.
More recently, machine learning techniques, particularly neural conditional density estimators (NCDEs), have provided powerful methods for approximating likelihood functions, posterior distributions and the likelihood ratio \citep{cranmer_frontier_2020}.

In complex applications, small-scale contaminations or unmodelled phenomena make it impractical to specify every detail of the DGP, particularly with large datasets \citep{miller_robust_2019}. Although practitioners are aware of the gap between simulations and reality, inference often proceeds as if the model is perfectly specified. This assumption is commonly violated when the true distribution lies outside the set of models considered $P_\star \notin \mathcal{P}$ (the $\mathcal{M}$-open scenario), in contrast to $\mathcal{M}$-closed settings where $P_\star \in \mathcal{P}$ \citep{bernardo_bayesian_2009, le_bayes_2017, yao_using_2018}.

The primary goal of misspecification-robust methods is to deliver reliable and useful inference even when in the $\mathcal{M}$-open scenario. We define robustness in the classical sense of \citet{huber_robust_2009}: small deviations from the assumed model should yield only small changes in inference. This ensures that modest discrepancies between the assumed model and the true distribution do not disproportionately affect our conclusions.

Model misspecification poses a major challenge in SBI, leading to empirically observed unreliable inferences \citep{cannon_investigating_2022, schmitt_detecting_2024} and violating typical theoretical assumptions, which demands separate treatment \citep{legramanti_concentration_2025, marin_relevant_2014, frazier_model_2020}. Although recently there has been considerable work on misspecification-robust methods in SBI across different areas of statistics \citep{bharti_approximate_2022, dellaporta_robust_2022, frazier_robust_2021} and machine learning \citep{huang_learning_2023, kelly_misspecification-robust_2024, ward_robust_2022}, no comprehensive work has tied them all together. This paper provides a thorough review and synthesis of research on model misspecification within an SBI context. We describe the issue of SBI with misspecified models, explore how various SBI methods are affected by this issue, and collate recent strategies for enhancing the robustness of these methods. In doing so, we aim to bring clarity to a dispersed literature, offer practical insights for real-world applications, and outline promising directions for future work.

In Section~\ref{sec:sbi_mm}, we examine three primary SBI approaches—approximate Bayesian computation (ABC), Bayesian synthetic likelihood (BSL), and neural conditional density estimation (NCDE)—and describe and demonstrate how each method is susceptible to model misspecification. In Section~\ref{sec:methods}, we categorise recent methods into three robust strategies for handling model misspecification in SBI: robust summary statistics, generalised Bayesian inference, and error modelling and adjustment parameters. We then illustrate the application of these strategies through a running example in Section~\ref{sec:toy_robust}. Finally, in Section~\ref{sec:discussion}, we conclude with a discussion and outline possible future directions for addressing model misspecification in SBI.

\section{SBI and model misspecification}\label{sec:sbi_mm}
In this section, we begin by introducing the necessary background on Bayesian inference, including a general discussion of model misspecification. We then provide an overview of simulation-based inference, focusing on three key approaches: approximate Bayesian computation, Bayesian synthetic likelihood, and neural conditional density estimation. Next, we examine model misspecification within the SBI framework, with an emphasis on theoretical insights. Finally, we present an illustrative example—a misspecified MA(1) model—to demonstrate how standard SBI methods can be adversely affected by model misspecification.

\subsection{Bayesian inference preliminaries}
Bayesian inference quantifies uncertainty in our model parameters by updating our initial beliefs using observations. 
The target of Bayesian inference for data $\mathbf{y}$ and model parameters $\bm{\theta}$ is the posterior distribution,
\begin{equation*}
    \pi(\bm{\theta} \mid \bm{y}) = \frac{p(\bm{y} \mid \bm{\theta}) \pi(\bm{\theta})}{\int_{\Theta} p(\bm{y} \mid \bm{\theta}) \pi(\bm{\theta}) \, \mathrm{d}\bm{\theta}},
\end{equation*}
where $\pi(\boldsymbol{\theta})$ is the density of the prior distribution, which represents the modeller's initial beliefs, and $p(\mathbf{y} \mid \boldsymbol{\theta})$ is the likelihood function, and $\Theta$ is the model parameter space. In practice, we often work with the joint distribution, which is proportional to the posterior up to a normalising constant,
\begin{equation*}\label{eq:joint_bayesian}
    \pi(\bm{\theta} \mid \bm{y}) \propto  \pi(\bm{\theta}, \bm{y}) = p(\bm{y} \mid \bm{\theta}) \pi(\bm{\theta}).
\end{equation*}
The DGP couples the natural phenomenon itself with data collection and processing, to produce observational data. We denote the observed data as $\mathbf{y} = (y_1, \ldots, y_n)^\top \in \mathcal{Y}$, which has been generated by the true DGP, $P_\star^{(n)}$ (hereafter omitting the $n$ superscript for notational brevity).
Of course, we do not have access to $P_\star$. Instead, we can specify an assumed DGP with a class of parametric models, $P_{\bm{\theta}}$.
Let $\mathcal{P} = \{ P_{\bm{\theta}} \colon \bm{\theta} \in \Theta \}$ be the family of candidate distributions.
We write $p_{\star}(\bm{y})$ for the density of $P_{\star}$, and $p(\bm{y} \mid \bm{\theta})$ for the density of $P_{\bm{\theta}}$, which represents the likelihood when $\bm{y}$ is fixed and $\bm{\theta}$ is varied. 

In Bayesian inference, model misspecification can arise from two interrelated components: the DGP and the prior distribution. While we focus on misspecification of the DGP in this paper, another form of misspecification arises when the information in the prior and likelihood are in conflict, and poor inferences can result from this even with a correctly specified DGP \citep{evans_checking_2006}. See \citet{chakraborty_weakly_2023} for a discussion of prior-data conflict checking in likelihood-free inference.

Rather than striving for a perfectly ``true'' model—which is often unattainable in practice—we treat model building as an iterative process. We begin with simpler models and only add complexity when it provides clearer insights or better fit, following George Box’s adage that ``all models are wrong, but some are useful'' \citep{box_science_1976, box_sampling_1980}. This approach underpins a principled Bayesian workflow \citep{betancourt_towards_2020, gabry_visualization_2019, gelman_bayesian_2020, schad_toward_2021}, which includes model building, inference, model checking, evaluation, and expansion.
Assessing model fit through computational validation and model evaluation techniques is a key part of this workflow, helping to identify potential misspecifications. For instance, posterior predictive checks assess whether the model can reproduce essential features of the observed data \citep{gelman_philosophy_2013}.
When checks indicate poor fit, the modeller can refine the model through approaches such as revising priors, incorporating additional data, reconsidering key assumptions, or expanding the model structure. However, indiscriminately increasing model complexity can obscure insights and hinder meaningful scientific inference \citep{mcelreath_statistical_2018, miller_robust_2019}.
While the Bayesian framework can, in principle, reconcile model complexity and data fit through Occam’s razor—via the model evidence \citep[see][Chapter 28]{mackay_information_2003}—this approach becomes problematic for SBI where only insufficient summary statistics are available \citep{robert_lack_2011, marin_relevant_2014}.

When we are in the $\mathcal{M}$-closed setting, the Bayesian approach  (with appropriate priors and sufficient computational power) is justified as the uniquely optimal method for tasks such as decision-making \citep{savage_foundations_1954} and information processing \citep{zellner_optimal_1988}. Moreover, it is ``consistent'', meaning that as more data accumulates, the posterior distribution concentrates on the true parameter value.

In contrast, in the $\mathcal{M}$-open scenario, under regularity conditions, standard Bayesian inference will concentrate around the pseudo-true parameter,
\begin{equation*}
    \bm{\theta}_{\star} = \argmin_{\bm{\theta} \in \Theta} D_{\mathrm{KL}}(P_{\star} \| P_{\bm{\theta}}) =  \argmin_{\bm{\theta} \in \Theta}  \int_{\mathcal{Y}} p_{\star}(\bm{y}) \log \left( \frac{p_{\star}(\bm{y})}{p(\bm{y} \mid \bm{\theta})} \right) \, \mathrm{d}\bm{y}
\end{equation*}
i.e.\ the value that minimises the Kullback-Leibler divergence between the true model and the assumed model \citep{berk_limiting_1966, walker_bayesian_2013}. Since
\begin{equation*}
     \int_{\mathcal{Y}} p_{\star}(\bm{y}) \log \left( \frac{p_{\star}(\bm{y})}{p(\bm{y} \mid \bm{\theta})} \right) \, \mathrm{d}\bm{y} = - \int_{\mathcal{Y}} p_{\star}(\bm{y}) \log p(\bm{y} \mid \bm{\theta}) \, \mathrm{d}\bm{y} + \text{c},
\end{equation*}
where $c$ is a constant with respect to $\bm{\theta}$, the same parameter $\bm{\theta}_{\star}$, equivalently emerges from maximising the expected log-likelihood $\mathbb{E}_{p_{\star}} [ \log p(\bm{y} \mid \bm{\theta})]$. By the likelihood principle, Bayesian updating is primarily driven by the likelihood, so as the prior influence diminishes with more data, the posterior concentrates around parameters that maximise the expected log-likelihood. Hence, standard Bayesian updating concentrates around the model point(s) that is statistically closest to $P_{\star}$ in KL terms.
Nonetheless, the resulting pseudo-true parameter(s) may not always be meaningful or desirable, particularly in the presence of contaminated data or outliers, which the KL divergence is particularly sensitive to \citep{basu_robust_1998, jewson_principles_2018}. For a more extensive discussion of the challenges and new developments in Bayesian inference under model misspecification, see the recent review papers by \citet{burkner_models_2023} and \citet{nott_bayesian_2024}.

\subsection{Background on SBI}
We are interested in conducting Bayesian inference when the likelihood function is unavailable or intractable, but we can readily simulate data from the model. 
We focus on approximate Bayesian computation (ABC), Bayesian synthetic likelihood (BSL), and neural conditional density estimators (NCDE), as the model misspecification developments in SBI have been based predominantly on these methods. Nonetheless, alternative strategies such as non-neural conditional density estimators \citep{forbes_summary_2022, haggstrom_fast_2024} and density-ratio estimation \citep{hermans_likelihood-free_2020, thomas_likelihood-free_2022} are also available. In addition, frequentist approaches \citep{cranmer_approximating_2016, dalmasso_likelihood-free_2024, warne_generalised_2023} offers another perspective, but our review remains focused on Bayesian methods.

In practice, SBI methods often reduce computational burden by mapping the full dataset to a set of summary statistics.
Let $S \colon \R^n \rightarrow \R^d$ be the summary statistic mapping, where $n$ is the size of the dataset, $d$ is the number of summaries, and $n \geq d$ and $d \geq d_{\bm{\theta}}$ where $d_{\bm{\theta}}$ denotes the number of model parameters.
Ideally, such a mapping would be sufficient, meaning that there is no information loss. However, in practice, our summary mappings are often insufficient—that is, they result in information loss. The choice of the number of summaries represents a trade-off between retaining information and reducing computational cost \citep{blum_comparative_2013}.
Although this paper primarily focuses on inference with summary statistics, the question of whether or not to summarise data remains an ongoing question in SBI \citep{chen_is_2023, drovandi_comparison_2022}. Further, increasingly automated methods are used for constructing summaries, especially in neural network–based approaches \citep{albert_learning_2022, chen_neural_2021, fearnhead_constructing_2012, jiang_learning_2017}.


\subsubsection{Approximate Bayesian computation.}
ABC has traditionally been the most popular class of method in SBI \citep{beaumont_approximate_2010, lintusaari_fundamentals_2017, martin_approximating_2024, pesonen_abc_2023, sisson_handbook_2018,  toni_approximate_2009, warne_simulation_2019}. 
The core idea of ABC is to generate simulated data $\bm{x}$ for a given parameter value $\bm{\theta}^*$ and build an approximate posterior by retaining all $\bm{\theta}^*$ that produce pseudo-data ``close enough'' to the observed data $\bm{y}$. This closeness is determined using a discrepancy function $\rho$ and a kernel function $K_{\epsilon}(\cdot)$ that allocates higher weight to lower discrepancies.
Because ABC relies only on generating data from a model rather than specifying a parametric likelihood function, it is often viewed as a nonparametric approach to inference \citep{blum_approximate_2010}.

 
The simplest and original ABC algorithm, rejection ABC, directly draws parameter values from the prior and retains those producing simulations within a tolerance $\epsilon$, as detailed in Algorithm~\ref{alg:rej_abc} \citep{tavare_inferring_1997, pritchard_population_1999}.  
However, if the prior and posterior distributions differ significantly, rejection ABC can result in many wasted simulations associated with parameter values that are rejected. To address this inefficiency, more advanced schemes have been developed that generate simulations using more promising proposal parameter values based on Markov chain Monte Carlo (MCMC) methods \citep{marjoram_markov_2003} and sequential Monte Carlo (SMC) methods \citep{drovandi_estimation_2011, jasra_multilevel_2019, sisson_sequential_2007, warne_multilevel_2018}. 

\begin{algorithm}
\caption{Rejection ABC}
\label{alg:rej_abc}
\begin{algorithmic}[1]
\ENSURE Prior distribution $\pi(\bm{\theta})$, tolerance $\epsilon$, number of samples $N$
\REQUIRE ABC posterior samples
\STATE Initialise $\mathcal{A} = \varnothing$ \COMMENT{Set to store accepted parameter values}
\FOR{$i=1$ to $N$}
    \STATE Draw $\bm{\theta}^{(i)}$ from $\pi(\bm{\theta})$
    \STATE Simulate $\bm{x}^{(i)}$ from $P_{\bm{\theta}^{(i)}}$
    \STATE Compute discrepancy $\rho(S(\bm{x}^{(i)}), S(\bm{y}))$
    \IF{$\rho(S(\bm{x}^{(i)}), S(\bm{y})) \leq \epsilon$}
        \STATE Accept $\bm{\theta}^{(i)}$ and add $\bm{\theta}^{(i)}$ to $\mathcal{A}$
    \ENDIF
\ENDFOR
\RETURN $\mathcal{A}$ \COMMENT{Return the set of accepted parameters values}
\end{algorithmic}
\end{algorithm}

Despite these improvements, all ABC methods face challenges due to the ``curse of dimensionality'' when dealing with high-dimensional summary statistics \citep{barber_rate_2015, csillery_abc_2012}. As dimensionality increases, either the tolerance for closeness needs to be relaxed, leading to increased approximation error, or fewer simulations are accepted, requiring a larger number of simulations to maintain a given number of posterior samples and thereby inflating computational costs. 

\subsubsection{Bayesian synthetic likelihood.}
Unlike ABC, which relies on comparing simulated data to observed data through a discrepancy function, BSL uses a parametric estimator to approximate the intractable likelihood function. The synthetic likelihood approach was first proposed by \citet{wood_statistical_2010}, who employed a multivariate normal distribution to approximate the likelihood of the summary statistics. This normality assumption could be motivated by the central limit theorem when the summary statistics are sums or averages of a large number of independent random variables.

\citet{price_bayesian_2018} incorporated the synthetic likelihood into a Bayesian framework, leading to the development of BSL. In standard BSL, the synthetic likelihood function is defined as: 
\begin{equation*}
g(S(\bm{y}) \mid \bm{\theta}) = \mathcal{N}(S(\bm{y}) \mid \mu(\bm{\theta}), \Sigma(\bm{\theta})),
\end{equation*}
where $\mu(\bm{\theta})$ and $\Sigma(\bm{\theta})$ are the mean and covariance matrix of the summary statistics at parameter value $\bm{\theta}$. 
Typically, $\mu(\bm{\theta})$ and $\Sigma(\bm{\theta})$ are unknown, and we use estimates $\widehat{\mu}(\bm{\theta})$ and $\widehat{\Sigma}(\bm{\theta})$ estimated using $m$ independent simulations. 
Consequently, each synthetic likelihood evaluation requires generating $m$ pseudo-datasets. 
The synthetic likelihood can be integrated into a Metropolis-Hastings MCMC algorithm to sample from the approximate posterior distribution \citep{price_bayesian_2018} or used within variational Bayes methods, such as variational BSL \citep{ong_variational_2018}.

A drawback of BSL is the simulation cost to attain the estimates $\widehat{\mu}(\bm{\theta})$ and $\widehat{\Sigma}(\bm{\theta})$ at each likelihood evaluation. If the covariance is assumed to be dense, the simulation cost scales quadratically in the dimension of the summaries \citep{priddle_efficient_2022}. Nevertheless, BSL can scale more favourably than ABC in higher-dimensional settings \citep{frazier_bayesian_2023}.

\subsubsection{Neural conditional density estimation.} 
NCDE approaches to SBI approximate the intractable conditional density of interest by learning a surrogate density from model simulations using neural networks. These methods have gained popularity due to their ability to handle high-dimensional data and parameters effectively. This is in contrast to traditional density estimation approaches, such as kernel density estimation (KDE) \citep{rosenblatt_remarks_1956, parzen_estimation_1962}, which suffer from the curse of dimensionality and become impractical in higher-dimensional settings.


While we use notation specific to posterior distributions, since this is the most common use case for CDE in SBI, the definitions and approaches we present apply to any conditional probability distribution.
In SBI with data summarisation, one typically obtains samples from the joint distribution $\pi(\bm{\theta}, S(\bm{y}))$. 
Amortised inference proceeds by learning a single conditional density $\pi(\bm{\theta} \mid S(\bm{x}))$ that can be evaluated at any new summary $S(\bm{x}) \in \mathcal{S}$, thus avoiding repeated, computationally expensive inference runs \citep[see][for a review]{zammit-mangion_neural_2024}.
In practice, this amounts to constructing a bivariate mapping  $\pi(\cdot \mid \cdot) \colon \Theta \times \mathcal{S} \rightarrow [0, \infty)$, where for each $\bm{\theta} \in \Theta$ and $S(\bm{x}) \in \mathcal{S}$ we can directly evaluate $\pi(\bm{\theta} \mid S(\bm{x}))$. 
Ideally the function class $\mathcal{Q}$ should be expressive enough to approximate $\pi(\cdot \mid S(\cdot))$ arbitrarily well, ensuring an accurate conditional probability density function for any pair $(\bm{\theta}, S(\bm{x}))$.
Neural networks have emerged as a powerful tool in this context, providing the necessary flexibility and capacity to model complex conditional densities.
In particular, normalising flows have been employed for density estimation due to their expressiveness, computational efficiency in density evaluation and sampling, and ease of training \citep{rezende_variational_2015}.
Subsequent advancements have enhanced the capabilities of normalising flows \citep{durkan_neural_2019, papamakarios_normalizing_2021}, making them highly effective for modelling complex distributions.
This has led to a surge in the popularity of NCDE methods within SBI.

Most commonly, NCDE approaches target the posterior distribution directly.
When neural networks are used for this posterior approximation, the method is referred to as neural posterior estimation (NPE) \citep{papamakarios_fast_2016,  lueckmann_flexible_2017, greenberg_automatic_2019}.
The aim is to learn a neural network approximation $\widehat{q}_{\bm{\phi}}(\cdot \mid \cdot)$ that minimises the forward KL,
\begin{equation}\label{eq:npe_loss}
    \bm{\phi}^* = \argmin_{\phi \in \Phi} \mathbb{E}_{\bm{\theta} \sim \pi(\bm{\theta}), \, \bm{x} \sim P_{\bm{\theta}}} [- \log q_{\phi} (\bm{\theta} \mid S(\bm{x}))],
\end{equation}
where $\bm{\phi}$ are the neural network parameters and $\bm{\phi}^*$ denotes the optimised values used during inference. 
We approximate this expectation using Monte Carlo estimation based on our simulated training dataset $\mathcal{D} = \{(\bm{\theta}^{(i)}, S(\bm{x}^{(i)})) \}_{i=1}^m$.
This procedure is detailed in Algorithm~\ref{alg:npe}.


\begin{algorithm}
\caption{Neural posterior estimation (NPE)}
\label{alg:npe}
\begin{algorithmic}[1]
\ENSURE Prior distribution $\pi(\bm{\theta})$, number of simulations $m$
\REQUIRE Neural network approximation $\widehat q_{\bm{\phi^*}}(\cdot \mid S)$
    \STATE Initialise dataset $\mathcal{D} = \varnothing$
    \FOR{$i=1$ to $m$}
        \STATE Sample parameter $\bm{\theta}^{(i)} \sim \pi(\bm{\theta})$
        \STATE Simulate data $\bm{x}^{(i)} \sim P_{\bm{\theta}^{(i)}}$
        \STATE Compute summary statistics $S(\bm{x}^{(i)})$
        \STATE Add $(\bm{\theta}^{(i)}, S(\bm{x}^{(i)}))$ to $\mathcal{D}$
    \ENDFOR
       \STATE Train $\widehat q_{\bm{\phi}}(\cdot \mid S)$ on $\mathcal{D}$ using Equation~\ref{eq:npe_loss}

    \RETURN $\widehat q_{\bm{\phi^*}}(\cdot \mid S)$
\end{algorithmic}
\end{algorithm}

If the amortised approach offers little benefit, or if an accurate NCDE cannot be efficiently learned from prior predictive simulations, one may employ sequential sampling schemes that run simulations in rounds \citep{papamakarios_fast_2016}. This strategy generates more simulations in regions of interest, with the intent of improving the efficiency and accuracy of the inference for the given observed data. When applied to NPE, this is termed sequential neural posterior estimation (SNPE). Another strategy to focus simulations in regions of interest is to first obtain samples via ABC methods to create a training dataset that more closely resembles the observed data, termed pre-conditioned NPE, as proposed by \citet{wang_preconditioned_2024}.

Another set of methods focuses on directly estimating the likelihood function itself \citep{cranmer_unifying_2016, lueckmann_likelihood-free_2019, papamakarios_sequential_2019}. These approaches learn the kernel $\pi(S(\cdot) \mid \cdot)$,
and then by holding $\bm{x}$ fixed and allowing $\bm{\theta}$ to vary, we obtain a likelihood surrogate $q_{\phi}(S(\bm{x}) \mid \bm{\theta})$. This technique is referred to as neural likelihood estimation (NLE), with the sequential version known as SNLE \citep{papamakarios_sequential_2019}. With a tractable density estimate of the likelihood, we can then obtain posterior samples using standard techniques such as MCMC. 

ABC methods have also incorporated elements of CDE and parametric approaches. For instance, \citet{blum_non-linear_2010} proposed using non-linear regression models, such as neural networks, to estimate the relationship between summary statistics and parameters in ABC. This approach serves as a precursor to modern neural methods, linking traditional ABC techniques—like linear regression adjustment in \citet{beaumont_approximate_2002}—to neural network-based approaches. Other methods, such as regression density estimation \citep{fan_approximate_2013}, learn the distribution of $S \mid \bm{\theta}$ within the ABC framework. 

Similarly, Gaussian process (GP) surrogate models have been employed in ABC to model the discrepancy between simulated and observed data efficiently, and to approximate the synthetic likelihood. For example, Bayesian optimisation for likelihood-free inference (BOLFI) \citep{gutmann_bayesian_2016} uses GPs to model the discrepancy function, guiding simulations toward informative regions of the parameter space and significantly reducing the number of required simulations.
Further work use Bayesian optimisation and decision making under uncertainty to achieve accurate, efficient posterior estimates \citep{jarvenpaa_efficient_2019, jarvenpaa_parallel_2021, oliveira_no-regret_2021}.


\subsection{Model misspecification in SBI}\label{sec:mm_sbi}
SBI methods often rely on summary statistics rather than directly using the full data, primarily due to computational constraints and high data dimensionality. Since standard notions of misspecification typically hinge on the full data likelihood, we must adapt them for use with summary statistics.
Following \citet{marin_relevant_2014}, we define the binding functions\footnote{The term ``binding function'' originates from the indirect inference literature and refers to the mapping between parameters and expected summaries.} for the simulated summaries of the model, $b(\bm{\theta}) = \mathbb{E}[S(\bm{x})\mid\bm{\theta}]$ and similarly for the observed summaries $b_{\star} = \mathbb{E}[S(\bm{y})]$. A model is said to be \textit{incompatible} if
\begin{equation}\label{eq:incompatible}
    \epsilon^* = \inf_{\bm{\theta} \in \Theta} d(b_{\star}, b(\bm{\theta})) > 0,
\end{equation}
indicating that no parameter value within the model can perfectly replicate the expected model summaries with those of the observed data. Alternatively, the model is considered to be compatible if $\epsilon^* = 0$, meaning there exists at least one parameter value for which $b(\bm{\theta}) = b_{\star}$. This distinction between compatibility and incompatibility is important because, even when we are unable to exactly match the full observed data, compatible summaries can be achieved by carefully selecting or constructing summaries that the model represents well. 



ABC compares observed and simulated summaries using an absolute discrepancy.
Under model misspecification, the ABC posterior has been shown by \citet{frazier_model_2020} to concentrate around a pseudo-true parameter value,
\begin{equation*}\label{eq:abc_pseudotrue}
    \bm{\theta}_{\star} = \arginf_{\theta \in \Theta} d(b_{\star}, b(\bm{\theta})).
\end{equation*}
Thus, the ABC pseudo-true parameter depends on both the choice of summary statistics and the discrepancy function, highlighting the importance of carefully selecting these components. While exactly replicating the full dataset may be infeasible, using robust summaries (see Section~\ref{sec:robust_summaries}) can help achieve compatibility. Similarly, selecting a robust discrepancy (see Section~\ref{sec:gbi}) might lead to a more meaningful convergence of the ABC posterior under model misspecification.

While ABC exhibits some robustness to misspecification in that it converges to a pseudo-true parameter—which may or may not be practically useful—it generally does not provide valid frequentist coverage. This limitation is not unique to ABC: standard Bayesian inference can also fail to achieve nominal coverage under misspecification \citep{kleijn_bernstein-von-mises_2012}. Moreover, unlike the usual Bernstein–von-Mises setting, the limiting distribution of the ABC posterior need not be Gaussian \citep{frazier_model_2020}. Further, popular post-processing adjustments (e.g., local regression) \citep{beaumont_approximate_2002} can worsen inference when the model is misspecified, shifting the posterior away from the pseudo-true parameter.

BSL implicitly uses a relative discrepancy based on the assumption that the summary statistics follow a Gaussian distribution. 
When the data summaries deviate substantially from Gaussianity, the method becomes inherently misspecified. 
As demonstrated by \citet{frazier_synthetic_2024}, BSL's incompatibility criterion weights the difference $\{ b_{\star} - b(\bm{\theta}) \}$ by $\Sigma(\bm{\theta})^{-1}$, effectively scaling the mismatch relative to the summary statistic covariance rather than treating it as an absolute distance as in ABC.
Although BSL also converges around parameters that align model-based synthetic summaries with the observed summaries (in a Gaussian sense), the posterior may become multi-modal or exhibit non-Gaussian shapes when the model is incompatible with the observed data distribution \citep{frazier_synthetic_2024}.

Whereas ABC and BSL have benefited from rigorous analysis of their misspecified behaviour, NCDE approaches currently lack an equivalent level of theoretical clarity.
Recent theoretical work \citep{frazier_statistical_2024} has provided insights under a compatibility assumption (i.e.\ well-specified models), but similar work for the misspecified scenario remains unexplored. Nonetheless, empirical studies indicate that model misspecification can adversely affect NCDE methods \citep{cannon_investigating_2022, schmitt_detecting_2024}. Intuitively, this arises because the neural network is evaluated on data distributions different from those it encountered during training.

Such challenges for neural networks under misspecification are closely related to out-of-distribution (OOD) generalisation \citep{hendrycks_many_2021, hendrycks_baseline_2022, yang_openood_2022}, a central concern in deep learning. 
For instance, normalising flows, which are commonly used for neural SBI, are known to struggle with OOD data \citep{kirichenko_why_2020}.
In standard deep learning, one typically gauges model generalisation through test and validation sets drawn from real data that the network did not see during training. Good performance on this unseen real data boosts confidence in the model’s broader applicability. Naturally, there remain open questions about what real data are included in the validation set—covariate shift, prior shift, and concept shift can still arise \citep[see][Chapter ~19]{murphy_probabilistic_2023}. Yet these issues differ from the model misspecification problem considered here, where the assumed model does not capture the true data-generating process. In SBI, a low validation loss only confirms strong performance on data generated by the assumed simulator; it does not guarantee good performance if the observed data differs substantially from the assumed model.

Diagnosing model misspecification is crucial for obtaining reliable results and guiding model refinement in ABC-based approaches. \citet{frazier_model_2020} propose two diagnostics for ABC: one examines the acceptance probability decay across tolerances—where deviations from linearity indicate misspecification—and the other compares posterior expectations from different ABC algorithms, with inconsistencies suggesting model issues. Similarly, \citet{gutmann_likelihood-free_2018} introduces classification accuracy as a discrepancy function in ABC, framing the inference problem as a classification task between simulated and observed data. High classification accuracy signals that the model struggles to reproduce observed data, thus pointing to potential misspecification.

Posterior predictive checks based on summary statistics are widely used in SBI \citep{bertorelle_abc_2010, wang_comprehensive_2024}. By simulating data from the posterior predictive distribution and comparing summary statistics with those from observed data, one can assess how well the model recovers actual observations. If posterior predictive intervals fail to cover most of the observed data, it indicates either model misspecification or poor inference performance. Additionally, \citet{chakraborty_weakly_2023} adapt prior–data conflict checks to the SBI setting, helping to identify inconsistencies between the prior and observed data that may signal misspecification.

Various goodness-of-fit tests have been proposed to assess model specification in SBI. \citet{dalmasso_validation_2020} presents a framework combining local two-sample tests at fixed parameter values and a global goodness-of-fit test across simulation parameters to detect misspecification in surrogate models. \citet{ramirez-hassan_testing_2024} propose a test statistic that converges to a chi-squared distribution under the null hypothesis  $\epsilon^*=0$ (as previously defined in Equation~\ref{eq:incompatible}), enabling hypothesis testing for model misspecification. \citet{schmitt_detecting_2024} use the maximum mean discrepancy (MMD) metric to detect differences between observed and simulated data distributions, employing hypothesis tests based on critical MMD values estimated from simulations.

\subsection{Illustrative example: misspecified MA(1)}\label{sec:ill_example}

We consider as a running example the misspecified moving average (MA) of order 1 example, as outlined in \citet{frazier_robust_2021}.
Our aim is to demonstrate on this toy example how ABC, BSL, and NCDE react differently to model misspecification. We will revisit this example in Section~\ref{sec:toy_robust} to show how the robust approaches described in Section~\ref{sec:methods} can be used to obtain desirable inferences.

In this misspecified example, our assumed data-generating process (DGP) is an MA(1) model:
\begin{equation*}
y_t = w_t + \theta w_{t_1}, \quad \-1 \leq \theta \leq 1, \quad w_t \sim \mathcal{N}(0, 1).
\end{equation*}

\begin{figure}[!htpb]
    \centering
    \includegraphics[width=0.8\linewidth]{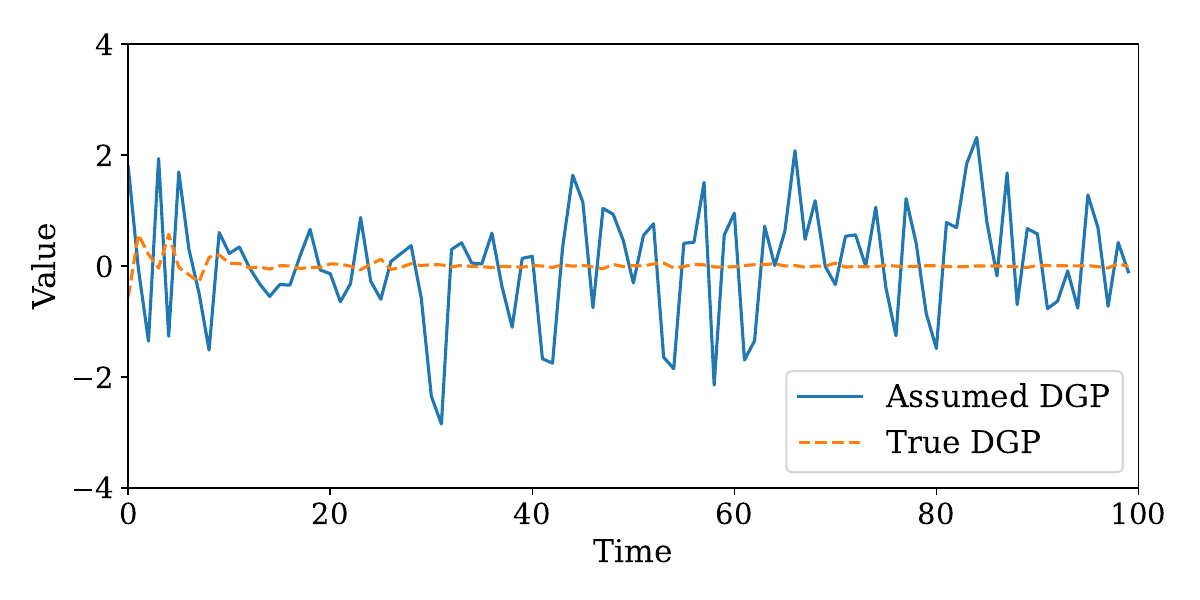}
    \caption{Comparison of full observed data (dashed line), and simulated data at the pseudo-true parameter $\theta=0$ (solid line).}
    \label{fig:dgp_comparison}
\end{figure}

However, the true DGP is a stochastic volatility model of the form:
\begin{equation*}
    y_t = \exp \left( \frac{z_t}{2} \right) u_t, \quad z_t = \omega + \kappa z_{t-1} + v_t + \sigma_v,
\end{equation*}
where $u_t$ and $v_t$ are both independent standard normal random variables, $\omega$ is a constant, $0 < \kappa$ and $\sigma_v < 1$.
We generate observed data with the ``true'' parameters $\omega= -0.76, \kappa = 0.90$ and $\sigma_v=0.36$. 
For the summaries, we take the two autocovariances $\zeta_j = \frac{1}{T} \sum_{i=j}^{T} x_i x_{i - j - 1}$, where $T=100$ is the number of observations, and $j \in \{0, 1 \}$.
We visualise both the assumed and true DGPs in Figure~\ref{fig:dgp_comparison}.
We set a uniform prior $\theta \sim \mathcal{U}(-1, 1)$.

Thus, there is a complete mismatch between the assumed and true DGPs. One might argue that Bayesian inference in such cases is ``meaningless—except perhaps as descriptive statistics'' \citep{freedman_so-called_2006}. However, in this example, there is still a parameter value around which we would hope the posterior would concentrate.
As we increase the number of observations of the assumed MA(1) model, the simulated summary statistics converge to $b(\theta)  = (1 + \theta^2, \theta)^\top$.
However, we can calculate that 
\begin{equation*}
    b_{\star} = (\exp \left( \frac{\omega}{1 - \rho} + \frac{\sigma_v^2}{2(1 - \rho^2)} \right) , 0)^{\top} \approx (0.0007, 0)^\top.
\end{equation*}
From both the full observed data and the observed summary statistics, it is clear there is no autocovariance.
We consider then the pseudo-true value of $\theta$ to be zero.
Further, it is at zero that the Euclidean distance between $b(\theta)$ and $b_{\star}$ is minimised. So not only does this pseudo-true make sense as the minimum distance between $b(\theta)$ and $b_{\star}$, it also has an interpretation that makes sense to the modeller, given that the observed autocovariance is near zero, we want $\theta$ values that produce autocovariances as close to zero as possible under the assumed model.

Of course, in practice, for such a simple example one would typically investigate further to develop a more accurate DGP. We use this example for pedagogical purposes: to show that even under misspecification where DGPs are fundamentally different, the posterior may concentrate on a parameter value that makes sense in the given situation and to illustrate the relevant diagnostics that a modeller could use to iterate on model building.
As a real-world analogue, the 2018 ``Volmageddon'' event provides an example in which volatility-linked financial products experienced catastrophic losses due to misspecified stochastic volatility models \citep{augustin_volmageddon_2021}. This failure underscores the danger of misspecified volatility assumptions. \citet{cannon_investigating_2022} emulated Volmageddon and found that common SBI methods yielded poor inference.

\begin{figure}[!htbp]
\centering
\begin{subfigure}[b]{0.49\textwidth}
    \includegraphics[width=\textwidth]{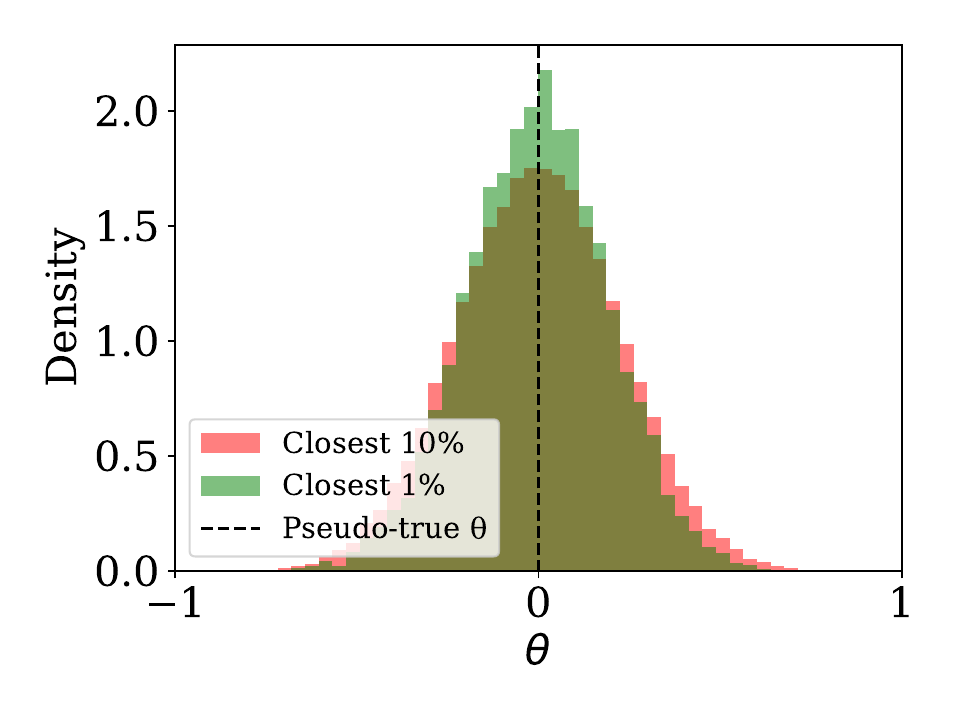}
    \label{fig:method1b}
\end{subfigure}
\hfill
\begin{subfigure}[b]{0.49\textwidth}
    \includegraphics[width=\textwidth]{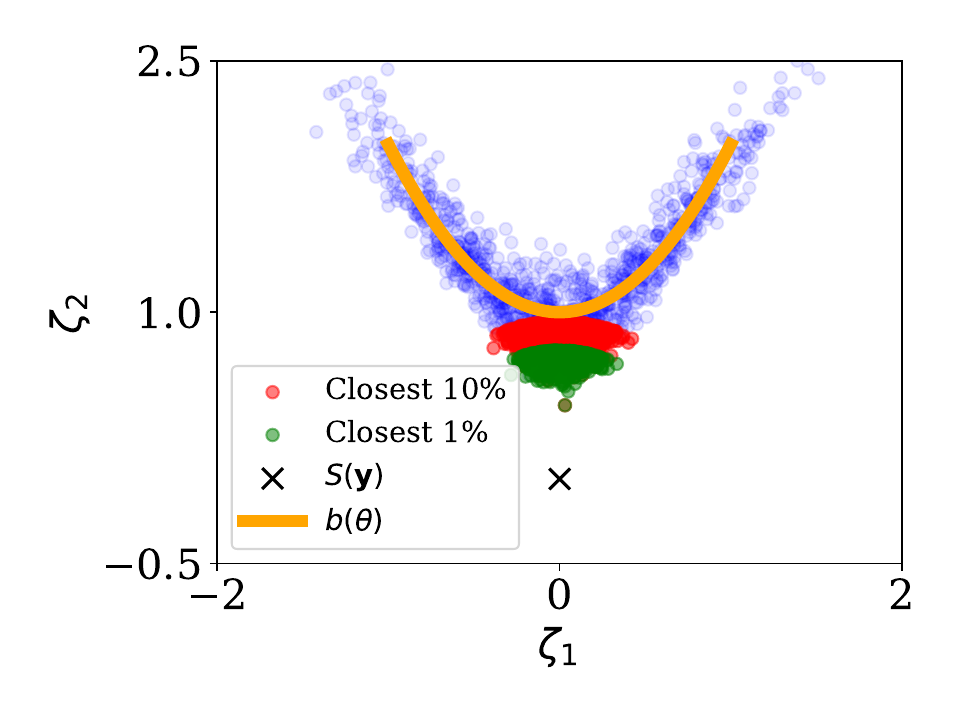}
    \label{fig:method1a}
\end{subfigure}
\vspace{-10mm}
\caption{Left: ABC posterior for $\theta$, with the pseudo-true value marked by a vertical dashed line at $\theta=0$. Right: Simulated summaries for the misspecified MA(1) example, showing observed summary ($\times$), simulated summaries (circles), and the binding function $b(\theta)$ (solid curve). }
\label{fig:misspec_ma1_abc}
\end{figure}

\begin{figure}[!htbp]
\centering
\begin{subfigure}[b]{0.49\textwidth}
    \includegraphics[width=\textwidth]{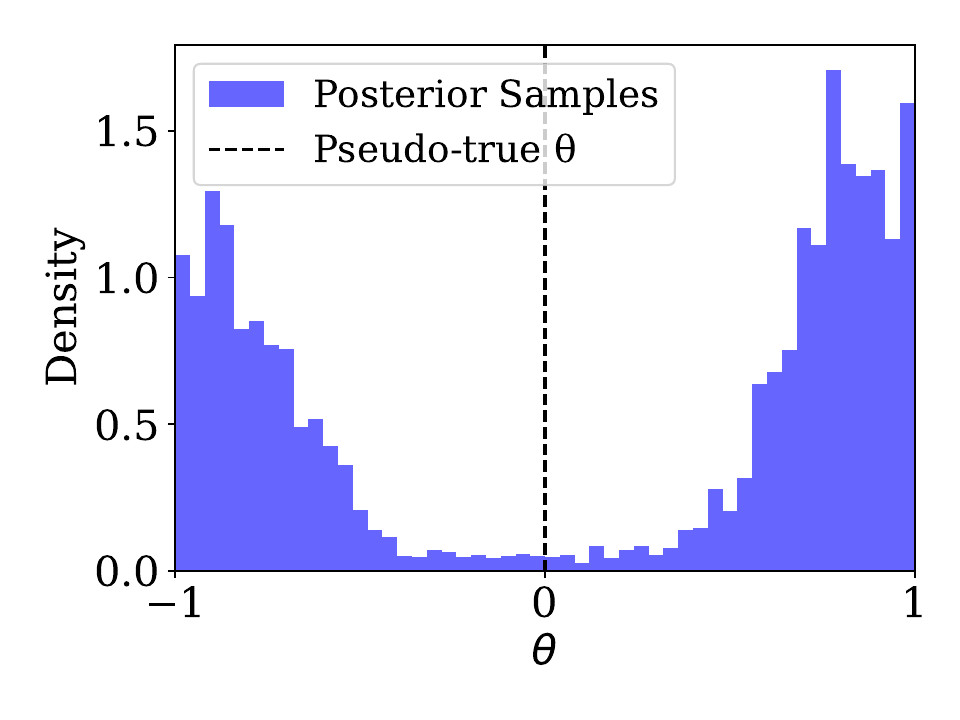}
    \label{fig:method2a}
\end{subfigure}
\hfill
\begin{subfigure}[b]{0.49\textwidth}
    \includegraphics[width=\textwidth]{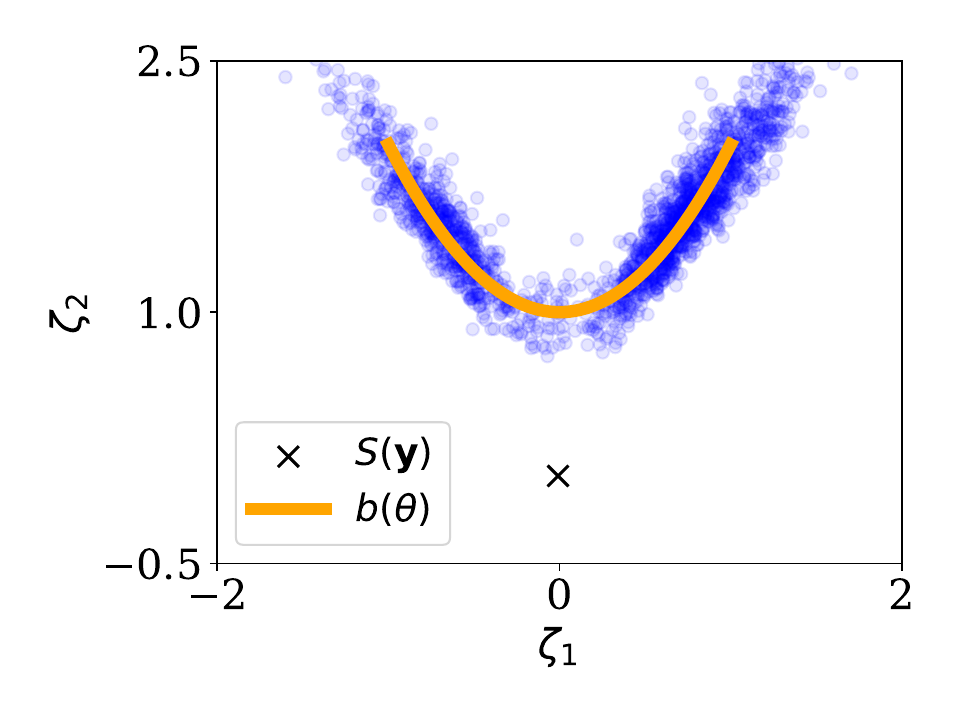}
    \label{fig:method2b}
\end{subfigure}
\vspace{-10mm}
\caption{Left: BSL posterior for $\theta$ for the misspecified MA(1) example. Right: Posterior predictive summaries (circles), observed summary ($\times$), and the binding function $b(\theta)$ (solid curve). 
}
\label{fig:misspec_ma1_bsl}
\end{figure}

\begin{figure}[!htbp]
\centering
\begin{subfigure}[b]{0.49\textwidth}
    \includegraphics[width=\textwidth]{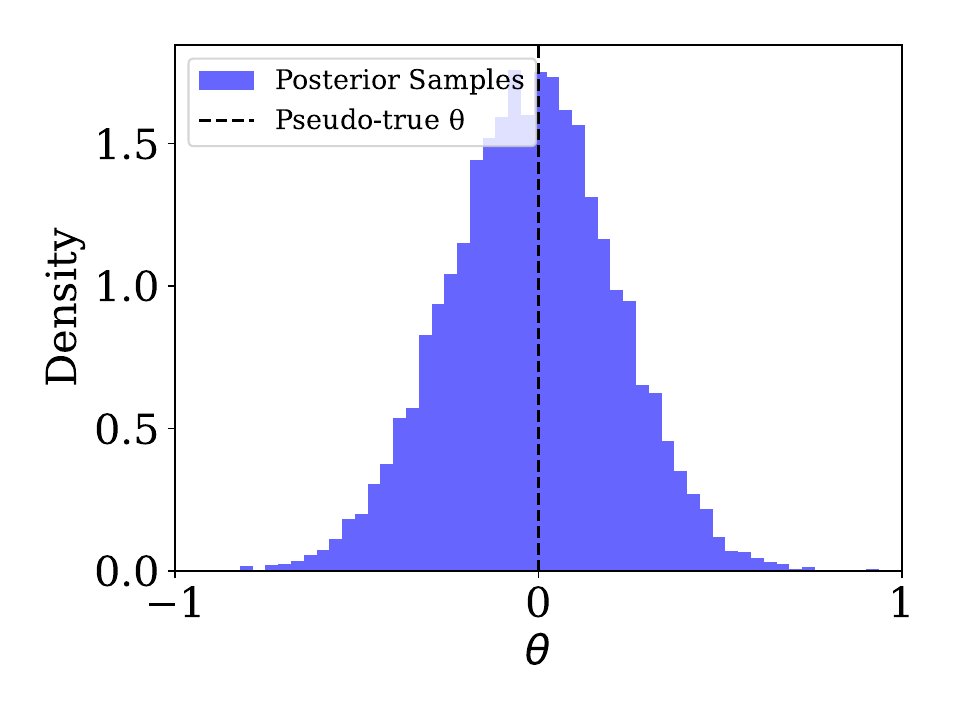}
    \label{fig:method3a_npe}
\end{subfigure}
\hfill
\begin{subfigure}[b]{0.49\textwidth}
    \includegraphics[width=\textwidth]{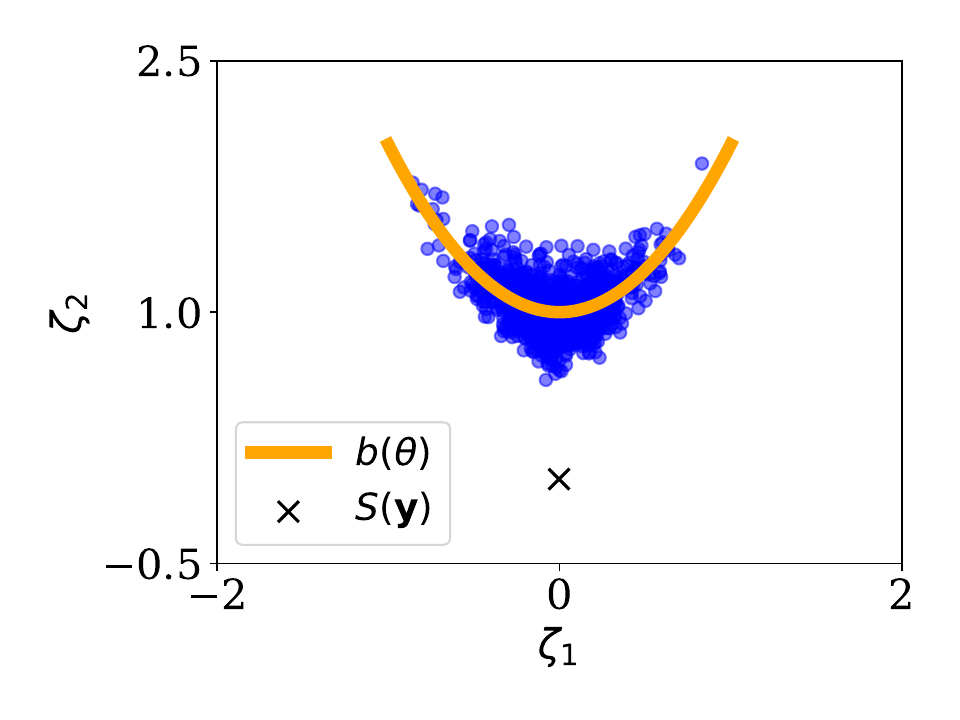}
    \label{fig:method3b_npe}
\end{subfigure} \\
\begin{subfigure}[b]{0.49\textwidth}
    \includegraphics[width=\textwidth]{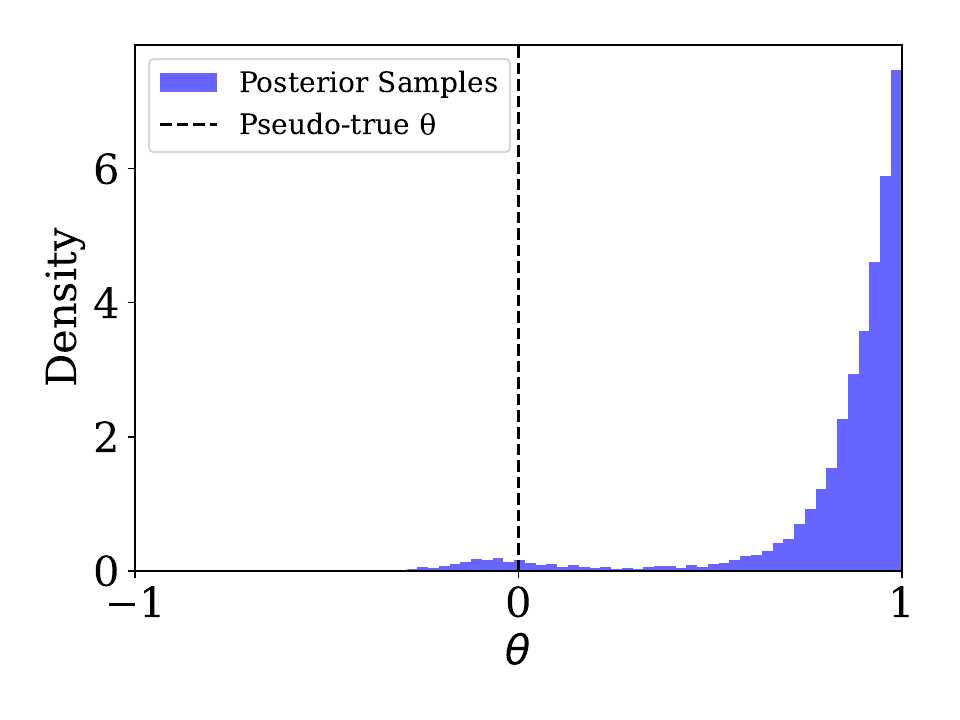}
    \label{fig:method3a_nle}
\end{subfigure}
\hfill
\begin{subfigure}[b]{0.49\textwidth}
    \includegraphics[width=\textwidth]{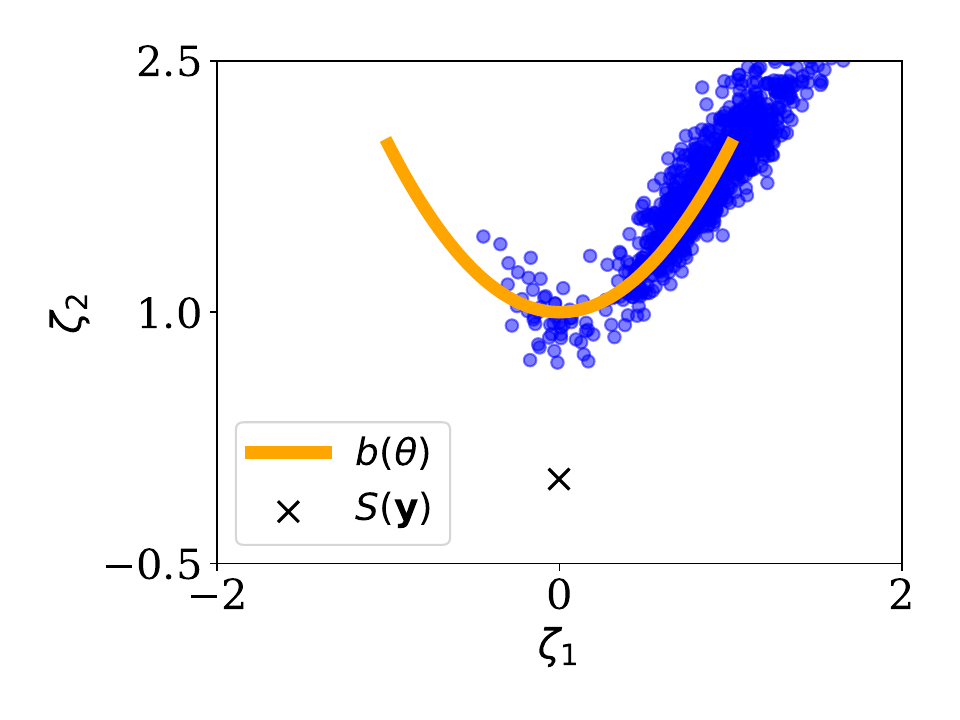}
    \label{fig:method3b_nle}
\end{subfigure}

\caption{Top row: NPE results. Left: Posterior for $\theta$, with the pseudo-true value at $\theta = 0$ indicated by a vertical dashed line. Right: Posterior predictive summaries (circles), observed summary ($\times$), and the binding function $b(\theta)$ (solid curve). Bottom row: Corresponding NLE results, with posterior on the left and posterior predictive checks on the right.}
\label{fig:misspec_ma1_cde}
\end{figure}

In this simple example, we can observe how each of the three broad categories of SBI methods reacts differently to model misspecification. 
In Figure~\ref{fig:misspec_ma1_abc}, we see some sense of an``inherent robustness'' in that ABC concentrates around a pseudo-true parameter value. In this instance, the pseudo-true parameter value defined in terms of the Euclidean norm is indeed zero which is the value we intuitively expect. As we decrease the ABC tolerance and appropriately scale up the number of simulations, we converge to the pseudo-true parameter value.
In Figure~\ref{fig:misspec_ma1_bsl}, the BSL posterior differs markedly from ABC, displaying multi-modality and concentrating away from the Euclidean-based pseudo-true value. This arises because BSL scales the mismatch $\{ b_{\star} - b(\bm{\theta}) \}$ by the covariance $\Sigma(\bm{\theta})$, treating it as a relative discrepancy. Multiple parameter values can then yield similar alignment in a Gaussian sense, producing a posterior that does not converge around $\theta = 0$ even though the observed autocovariance is near zero.
In Figure~\ref{fig:misspec_ma1_cde}, for this specific example, NPE performs well, whereas NLE concentrates on the boundary of the parameter space. Because the summaries (autocovariances of a moving average) are approximately Gaussian, NLE behaves similarly to BSL. It is challenging to \textit{a priori} predict how NPE and NLE will behave in such misspecified scenarios.


\FloatBarrier

\section{Robust approaches for SBI}\label{sec:methods}
We characterise three main strategies that have been used in the literature to attain misspecification-robust inference in SBI.
First, one can mitigate the misspecification of the full, complex data through the careful selection of robust summary statistics.
The second approach exploits generalised Bayesian inference, where robust inference can be achieved by using alternative loss functions. In particular, SBI can be carried out in the GBI framework through robust loss functions that are sample-based.
Third, directly modelling the discrepancy between the observed and simulated summary statistics through adjustment parameters can address data mismatches. 

\subsection{Robust summary statistics}\label{sec:robust_summaries}
When using summaries instead of the full dataset, model misspecification manifests as the inability to reproduce the observed summaries rather than the full observed data. 
By mapping the data to a carefully selected set of robust summaries, we can focus on the relevant aspects important for our inferential goals while mitigating the influence of complex noise or irrelevant features that may drive misspecification. The construction of summary statistics has received a great deal of attention in SBI (for a general reference \citealp[see][Chapter ~5]{sisson_handbook_2018}), but we limit our focus here to how we can construct summaries to be robust to model misspecification.

Adhering to a principled Bayesian workflow \citep{betancourt_towards_2020}, we distinguish between relevant and irrelevant model misfit. Instead of attempting to exactly match the true DGP with our assumed model, we concentrate on summarising the data into features, that are relevant for our analysis objectives, and disregard irrelevant features such as artifacts from data collection or processing. Non-robust inference can often be derailed by the influence of these irrelevant features.
In SBI, we aim to capture relevant features through the summary statistics. 
Ideally, we would like the constructed summary statistics selected to be robust, remaining reliable even when there are minor deviations from model assumptions.
As a common example, the median is a robust summary of central tendency due to its high breakdown point, meaning it can handle significant contamination without degrading the estimate.

When we conduct inference based on summary statistics, the inferential ideal is no longer the standard Bayesian posterior based on the full dataset but rather the partial posterior, $\pi(\bm{\theta} \mid S(\bm{y}))$, instead of the full posterior, $\pi(\bm{\theta} \mid \bm{y})$. \citet{doksum_consistent_1990} provide an example where the full posterior is inconsistent, but a partial posterior conditioned on appropriate summaries is consistent and asymptotically normal.
The partial posterior of interest may be based on sufficient summaries, where there is no information loss, or insufficient summaries, where information loss is incurred. Although inference based on summary statistics is often viewed as a necessary evil—since sufficient summaries are often not achievable and we lose information to gain computational efficiency—if the lost information pertains to irrelevant features, then using summaries may actually enhance robustness.


The Bayesian restricted likelihood approach proposed by \citet{lewis_bayesian_2021} maps the data to a set of insufficient summary statistics that are robust to certain data features, capturing aspects of interest while being insensitive to undesirable perturbations. For instance, M-estimators can reduce sensitivity to outliers (see \citealt[Chapter 5]{vaart_asymptotic_2000} for an overview). This was also recognised in \citet{ruli_robust_2020} who employ M-estimating functions to construct robust summary statistics for ABC. Recently, \citet{luciano_insufficient_2024} developed a Gibbs sampling method that conditions on robust insufficient summaries.

An alternative strategy is Bayesian data selection, which identifies parts of the data compatible with the assumed parametric model \citep{weinstein_bayesian_2023}. In this framework, the ``foreground'' is modelled with the parametric model and the ``background'' with a Bayesian nonparametric model. \citet{weinstein_bayesian_2023} propose a novel score to find lower-dimensional projections of the data effectively described by the parametric model foreground. This perspective is relevant for SBI, as these projections can serve as compatible summary statistics, allowing us to focus on data aspects that can be captured by our model simulations.

Also focused on the task of Bayesian data selection, \citet{huang_learning_2023} use the maximum mean discrepancy (MMD) in the loss function to robustly learn summaries for use in SBI.
Previous approaches to automated methods to construct summary statistics, while effective in capturing information, do not typically consider the scenario of model misspecification \citep{albert_learning_2022, chen_neural_2021, fearnhead_constructing_2012, jiang_learning_2017}. In contrast, \citet{huang_learning_2023} focus on learning informative summaries that are also robust to misspecification. 
They present two methods: first, embedding this within NPE and jointly learning both the neural network and the summary statistic network; second, using an autoencoder to learn robust summaries for ABC. For NPE, we now minimise the loss:
\begin{equation*}
        \argmin_{\bm{\phi} \in \Phi, \bm{\psi} \in \Psi} \mathbb{E}_{\bm{x}, \bm{\theta} \sim \pi(\bm{x}, \bm{\theta})} \left[- \log q_{\bm{\phi}} (\bm{\theta} \mid S_{\bm{\psi}}(\bm{x})) + \lambda \text{MMD}_k^2\big(S_{\psi}(\bm{x}),  S_{\bm{\psi}}(\bm{y})\big) \right],
\end{equation*}
where $S_{\bm{\psi}}(\cdot)$ is the summary mapping parametrised by $\bm{\psi}$, $k$ is the chosen kernel in MMD, and $\lambda$ is a hyperparameter that trades-off between efficiency and robustness. 

Additionally, \citet{bharti_approximate_2022} address model misspecification in ABC by incorporating domain experts in the summary selection process. Recognising the challenge of manually selecting informative summaries, they propose a sequential experimental design that actively involves the expert in a minimally intrusive manner. This approach also effectively handles misspecification by allowing the expert to identify and exclude misleading summaries.
Another strategy is modularised Bayesian inference which addresses model misspecification by decomposing the joint posterior into modules and using ``cutting feedback'' methods to selectively ignore misspecified modules \citep{bayarri_modularization_2009, yu_variational_2023}, which in our summary-based context means we can ignore the influence of summaries that adversely impact inference on some of the model parameters. \citet{chakraborty_modularized_2023} further extend this concept within SBI by applying a cutting feedback method based on a Gaussian mixture approximation of the joint posterior. 

\subsection{Generalised Bayesian inference}\label{sec:gbi}
Standard Bayesian inference can be highly sensitive to model misspecification because it effectively minimises the Kullback–Leibler (KL) divergence, which heavily penalises any small-probability mismatch between model and data \citep{basu_robust_1998, jewson_principles_2018}. To counteract this vulnerability, alternative loss functions can replace the usual log-likelihood update, forming the basis of generalised Bayesian inference (GBI). In this section, we explore how GBI ideas connect to SBI.

GBI extends the usual Bayesian update of beliefs \citep{bissiri_general_2016, knoblauch_optimization-centric_2022}. The generalised posterior (also known as the Gibbs posterior or pseudo-posterior) is defined as
\begin{equation}\label{eq:gbi}
\pi^{\mathsf{L}} (\bm{\theta} \mid \bm{y}) \propto \exp \left\{ -w \sum_{i=1}^{n} \mathsf{L}(y_i, \bm{\theta}) \right\} \pi(\bm{\theta}),
\end{equation}
where $\mathsf{L}(y_i, \bm{\theta}) \colon \mathcal{Y} \times \Theta \rightarrow \mathbb{R} $ is a loss function associated with the observed data $\bm{y}$, and $w$ is a calibration parameter. When $\mathsf{L}$ is the negative log-likelihood and $w=1$, we recover standard Bayesian inference. 
One can also derive \eqref{eq:gbi} through a variational perspective \citep{bissiri_general_2016, knoblauch_optimization-centric_2022}:
\begin{equation*}\label{eq:gbi_variational}
    \pi^{\mathsf{L}} (\bm{\theta} \mid \bm{y}) = \argmin_{q(\bm{\theta}) \in \Pi} \left\{ \sum_{i=1}^n \mathbb{E}_{q(\bm{\theta})} \left[ \mathsf{L}(y_i, \bm{\theta}) \right] + \text{KLD}(q(\bm{\theta}) || \pi(\bm{\theta})) \right\},
\end{equation*}
where $q(\bm{\theta})$ is any candidate distribution and $\Pi$ is the space of feasible solutions. We mainly focus here on addressing likelihood misspecification by choosing a robust or alternative $\mathsf{L}$. See \citet{knoblauch_optimization-centric_2022} for further generalisations involving different prior regularisation measures and spaces of feasible solutions.

Some loss functions are particularly suitable for models with intractable likelihoods because they require only the ability to generate samples. For a survey of sample-based distances, see \citet{bischoff_practical_2024}. For example, the kernel Stein discrepancy (KSD) \citep{chwialkowski_kernel_2016} has been used in GBI frameworks to handle likelihoods with intractable normalisation constants \citep{matsubara_robust_2022, matsubara_generalized_2023}. While KSD requires computation of the score function, the maximum mean discrepancy (MMD) \citep{gretton_kernel_2012} requires only samples and has been utilised in the MMD-Bayes method \citep{cherief-abdellatif_mmd-bayes_2020}, replacing the likelihood with an MMD loss. 

\subsubsection{ABC posterior is a generalised posterior.}\label{sec:abc_gbi}

The ABC posterior can be interpreted as a generalised posterior within the GBI framework \citep{miller_robust_2019, schmon_generalized_2020}. We can recast the usual ABC kernel $K_{\epsilon}\bigl(\rho(S(\bm{y}), S(\bm{x}))\bigr)$ directly as a loss \(\mathsf{L}(\bm{y},\bm{\theta})\) by integrating out \(\bm{x}\):
\begin{equation*}
  \mathsf{L}(\bm{y},\bm{\theta})
  \;=\;
  -\log 
  \int 
  K_{\epsilon}\!\bigl(\rho(S(\bm{y}), S(\bm{x}))\bigr)\, 
  p(\bm{x}\mid\bm{\theta})
  \,\mathrm{d}\bm{x}  \;\approx\; -\log \Bigl( \frac{1}{N} \sum_{i=1}^N K_{\epsilon}\!\bigl(\rho(S(\bm{y}), S(\bm{x_i}))\bigr) \Bigr),
\end{equation*}
where $\bm{x}_1,\dots,\bm{x}_N \sim P_{\bm{\theta}}$.
Exponentiating the Monte Carlo estimate of $\mathsf{L}(\bm{y},\bm{\theta})$ in the GBI update in Equation~\ref{eq:gbi} then yields the standard ABC weighting scheme based on \(\rho\) and \(\epsilon\).

\citet{miller_robust_2019} formalise a similar connection via ``coarsened posteriors,’’ which enhance robustness by conditioning on a neighbourhood of the empirical distribution rather than on the exact data. As a by-product of their work, they illustrate how ABC posteriors implicitly fit into a generalised posterior perspective through the kernel \(K_{\epsilon}\).

In their pioneering work, \citet{wilkinson_approximate_2013} recognise the possibility of model misspecification by treating discrepancies between the assumed and true DGPs as either model or measurement error. They show that if the actual model error is specified, ABC yields exact results under that assumption, reflecting how \(K_{\epsilon}\) could encode model error. 
Further, \citet{schmon_generalized_2020} interpret ABC’s accept/reject step as implicitly defining an error model. In practice, threshold-based or Gaussian kernels are chosen mostly for computational or heuristic reasons, but are likely to themselves be misspecified. For instance, rejection ABC can be viewed as exact inference with uniform model error on an \(\epsilon\)-ball around \(S(\bm{y})\), which is unlikely to reflect actual model error. To mitigate this potentially misspecified error assumption, one can adopt more flexible error distributions (see Section~\ref{sec:adjustment}). Moreover, since ABC can be cast as a GBI method, broader GBI strategies for robust inference naturally apply \citep{schmon_generalized_2020}.

There are two main ways by which generalised posteriors can address model misspecification: by adjusting the calibration parameter $w$ and by selecting a loss function $\mathsf{L}$ that is more robust to misspecification.

\subsubsection{Calibrating \texorpdfstring{$w$}{w}.} 
Reducing the weight on the negative log-likelihood (through $w < 1$) can limit the influence of a misspecified likelihood. 
This approach is referred to as tempered posteriors \citep{holmes_assigning_2017} or fractional posteriors \citep{bhattacharya_bayesian_2019}. One of the earliest methods on this is SafeBayes \citep{grunwald_safe_2012, grunwald_inconsistency_2017} which adaptively scales the likelihood to prevent inconsistency under misspecification, ensuring the posterior remains ``safe'' even when the model is not correctly specified. In \citet{miller_robust_2019}, they approximate their coarsened posterior, which is a generalised posterior, through tempering the likelihood.
While tempering may be a valid robust strategy for a genuine likelihood, tempering has generally proven ineffective for likelihood-based approaches that rely on simulated data. For example, \citet{frazier_synthetic_2024} consider tempering the synthetic likelihood, and \citet{gao_generalized_2023} consider tempering for neural likelihood approaches. In both cases, the strategy was found to be ineffective or even harmful. As demonstrated in \citet{frazier_synthetic_2024}, while tempering can alter the scale of the posterior approximation, it cannot change its mode or overall shape, and hence often fails to address the core issues arising from misspecification.


\subsubsection{Robust losses.} 
Generalised Bayesian inference (GBI) can be used to address model misspecification by choosing a robust loss function in Equation~\ref{eq:gbi}. As the ABC posterior is a generalised posterior, we can apply the same philosophy to motivate the choice of a robust distance. Relatedly, we have the result in \citet{frazier_model_2020} that the ABC distance determines the pseudo-true parameter, highlighting the importance of selecting distances that are less sensitive to minor model deviations. We will focus here on distance choices already considered for ABC. 




Integral probability metrics (IPMs) \citep{muller_integral_1997}, which include the MMD and Wasserstein distance, are a useful class of metrics for simulation-based inference (SBI). The general form of an IPM between two probability measures $\mathcal{P}$ and $\mathcal{Q}$ over a space $\mathcal{X}$ is defined as:
\begin{equation*}\label{eq:ipm}
\mathrm{IPM}_\mathcal{F}(P, Q) = \sup_{f \in \mathcal{F}} \left| \int_{\mathcal{X}} f(x) dP(x) - \int_{\mathcal{X}} f(x) \, \mathrm{d}Q(x) \right|,
\end{equation*}
where $\mathcal{F}$ is a class of functions.

Recent theoretical work by \citet{legramanti_concentration_2025} provides rigorous theory on the use of IPM-based distances in ABC. Their framework introduces the concept of Rademacher complexity to analyse the limiting properties of discrepancy-based ABC posteriors, including in non-i.i.d. and misspecified settings.

An early example of a robust distance in ABC is by \citet{park_k2-abc_2016} who introduce K2-ABC, which employs the MMD as the discrepancy function in ABC. An unbiased estimator of the squared MMD is given by:
\begin{equation*}\label{eq:unbiased_mmd}
    \widehat{\mathrm{MMD}}^2_u(P, Q) = \frac{1}{m(m-1)} \sum_{i \neq j} \mathcal{K}(\bm{x_i}, \bm{x_j}) + \frac{1}{n(n-1)} \sum_{i \neq j} \mathcal{K}(\bm{y_i}, \bm{y_j}) - \frac{2}{mn} \sum_{i,j} \mathcal{K}(\bm{x_i}, \bm{y_j}),
\end{equation*}
where $\{\bm{x_i}\}_{i=1}^m \sim P$,  $\{\bm{y_j}\}_{j=1}^n \sim Q$ and $\mathcal{K}$ is the chosen kernel, typically a Gaussian kernel. The ability to compute MMD solely based on samples makes it appealing for SBI.
MMD is a particularly attractive choice for achieving outlier robustness, as evident in its use for diagnosing model misspecification \citep{schmitt_detecting_2024} and learning robust summary statistics \citep{huang_learning_2023}.
Another application of MMD in SBI is the MMD posterior bootstrap method proposed by \citet{dellaporta_robust_2022}, which combines the posterior bootstrap with MMD estimators to achieve robust and highly parallelisable Bayesian inference.

Another distance choice investigated is the Wasserstein distance \citep{villani_optimal_2009}. \citet{bernton_approximate_2019} proposed using the Wasserstein distance between the empirical distributions of the observed and simulated data in ABC. They developed computational approximations to mitigate the super-quadratic scaling with the number of observations, making the approach more practical for large datasets.

Further divergences have been explored in ABC.
\citet{fujisawa_-abc_2021} utilised a $\gamma$-divergence estimator via a $k$-nearest-neighbor-based kernel density estimator. This approach exhibits the redescending property, automatically ignoring extreme outliers and enhancing robustness to heavy data contamination. \citet{jiang_approximate_2018} considered the KL divergence as the data discrepancy in ABC, aligning the ABC pseudo-true parameter with the traditional KL-based pseudo-true parameter in Bayesian inference \citep{muller_risk_2013}. \citet{frazier_robust_2020} investigated the use of the Hellinger and Cramér-von Mises (CvM) distances, demonstrating their potential for robustness under model misspecification. 

With the plethora of distance options, it is useful for modellers to know which are most promising for use in ABC. Of course, the most appropriate distance is problem-dependent, which is backed up by empirical results in \citet{drovandi_comparison_2022} who compared CvM, Wasserstein, and MMD distances in well-specified settings. In the misspecification setting, \citet{legramanti_concentration_2025} compared MMD, Wasserstein, and KL divergences within a misspecified Huber contamination model. They found that MMD performed best across various levels of misspecification, highlighting its effectiveness in robust inference.

Initially, the main motivation for investigating many choices of distance in ABC is their applicability both with and without summaries. For instance, the original motivation for K2-ABC in \citet{park_k2-abc_2016} was to avoid using insufficient summaries and made no mention of misspecification. However, it has fortuitously turned out that the use of MMD in ABC can be highly robust to model misspecification. This section applies to both summary-based and summary-free SBI.

In ABC, the loss used to define the generalised posterior is fairly explicit, as the modeller must select a distance and error kernel. For BSL and NCDE methods, an analogous loss function is also used — albeit less explicitly.
\citet{pacchiardi_generalized_2024} frame GBI using scoring rules \citep{gneiting_strictly_2007}, deriving the loss function from a proper scoring rule to measure the discrepancy between the model and the data. For instance, the BSL posterior corresponds to a generalised posterior with the Dawid–Sebastiani score \citep{dawid_coherent_1999}. 

NCDE methods typically aim to approximate the standard Bayesian posterior; for instance, NPE minimises the forward KL divergence as shown in Equation~\ref{eq:npe_loss}. As noted, the KL divergence is not robust, with the worst-case scenario being adversarial attacks—small, targeted perturbations that significantly affect the estimator's output. To enhance robustness against such adversarial perturbations, \citet{glockler_adversarial_2023} propose a regularisation scheme based on penalizing the Fisher information of the conditional density estimator.

Alternatively, we can train neural networks using loss functions other than the KL divergence to improve robustness under model misspecification. 
For example, \citet{gao_generalized_2023} introduce an amortised approach to GBI by training a neural network to predict the loss directly. Their method, amortised cost estimation (ACE), learns a surrogate model of the loss function, which eliminates the need for extensive simulations during inference. By employing robust loss functions, such as MMD, ACE can provide misspecification-robust inference.
The same principle applies to other SBI methods that rely on surrogate modelling of the loss function, such as BOLFI \citep{gutmann_bayesian_2016}, where the discrepancy being modelled can be selected to be robust.

\subsection{Error modelling and adjustment parameters}\label{sec:adjustment} 
Model misspecification often stems from structural discrepancies between the assumed model and the true DGP.
One general approach to address these discrepancies is to introduce an explicit error model, $p(\bm{y} \mid \bm{x})$, that represents how the observed data $\bm{y}$ deviate from the model’s simulations $\bm{x}$. Formally, we consider:
\begin{equation*}
    p(\bm{y} \mid \bm{\theta}) = \int  p(\bm{x} \mid \bm{\theta}) p(\bm{y} \mid \bm{x})  \, \mathrm{d}\bm{x},
\end{equation*}
where we assume $\bm{y}$ is conditionally independent of $\bm{\theta}$ given $\bm{x}$.
Under perfect model specification, this corresponds to using a Dirac measure $\delta(\bm{y} - \bm{x})$, which models there being no discrepancy.
However, as this assumption is often unrealistic, recent methods relax this assumption by specifying a non-trivial error model. For instance, this has been considered for SBI methods with \citet{bernaerts_combined_2023} adding Gaussian noise, while \citet{ward_robust_2022} use a spike-and-slab formulation.
Although identifying the true error distribution is challenging, even approximate choices can substantially improve robustness. This connects with the viewpoint that exactly specifying the model error through an ABC kernel leads to exact inference  \citep{wilkinson_approximate_2013} (see Section~\ref{sec:abc_gbi}).

A useful special case of these error models involves adjustment parameters, which directly shift the simulated outputs to better align with the observed data. 
When using summary statistics, the adjustment parameters can be motivated as shifting incompatible summaries.
Introducing additive adjustment parameters with the same dimension as the summary statistics $\bm{\Gamma} = (\gamma_1,\ldots,\gamma_d)^\top$,  for a given $\bm{\theta}$, we have: \begin{equation*}
S(\bm{y}) = S(\bm{x}) + \bm{\Gamma}.
\end{equation*}
With the inclusion of adjustment parameters, the likelihood in terms of the observed summary statistics becomes the convolution:
\begin{equation*}
g(S(\bm{y}) \mid \bm{\theta}) = \int g(S(\bm{x}) \mid \bm{\theta}) h(S(\bm{y}) - S(\bm{x})) \, \mathrm{d}\bm{x},
\end{equation*}
i.e., $p(\bm{y} \mid \bm{x}) = h(S(\bm{y}) - S(\bm{x})) = h(\bm{\Gamma})$, where $ h(\bm{\Gamma})$ is the density of $\bm{\Gamma}$. By explicitly inferring these adjustment parameters, the inference procedure ``absorbs'' model misspecification into these parameters, thus making the overall inference more robust.

Early work on deterministic computer models introduced adjustment parameters to account for the mismatch between the model and the observed data.
\citet{kennedy_bayesian_2001} model $\bm{\Gamma}$, termed the ``model inadequacy correction'', using a Gaussian process.
\citet{bayarri_modularization_2009} extend this approach within a modularised Bayesian framework and provide strategies for modelling $\bm{\Gamma}$.

In the context of SBI, an early method is the $\text{ABC}_{\mu}$ approach proposed by \citet{ratmann_model_2009}. 
Acknowledging the necessity of assessing model adequacy, \citet{ratmann_model_2009} introduce an unknown error term into the likelihood function, treating the ABC error tolerance as a random variable with its own exponential prior distribution.
These random variable ABC error tolerances, which are done for each summary, are the same as adjustment parameters $\bm{\Gamma}$, i.e., the discrepancy between simulated and observed summaries.

Sampling from the joint posterior distribution of both the model parameters and the error term allows for direct investigation of model misspecification through the posterior distribution of the error tolerance. This opportunity for model criticism is a common feature across all adjustment parameter methods. Under compatible summaries, as shown in \citet{frazier_robust_2021}, the posteriors for the components of $\bm{\Gamma}$ converge to their priors. Under model misspecification, the posterior of the adjustment parameters deviate from their priors to correct discrepancies between the model and the observed data. By examining which adjustment parameters differ significantly from their priors, we can identify incompatible summary statistics. If the summary statistics are carefully selected to be meaningful to a domain expert, this approach reveals precisely which aspects of the model are inadequate, facilitating model criticism and refinement.

\citet{frazier_robust_2021} proposed the robust BSL (RBSL) method, which introduces adjustment parameters to synthetic likelihood. In RBSL-M, the mean vector in the synthetic likelihood is adjusted:
\begin{equation*}\label{eq:rbslm}
    \mathcal{N}(S(\bm{y}) \mid \mu(\bm{\theta}) + \sigma(\bm{\theta}) \odot \bm{\Gamma}, \Sigma(\bm{\theta})),
\end{equation*}
where $\sigma(\bm{\theta}) = \text{diag}(\Sigma(\bm{\theta}))$ is the vector of standard deviations of the simulated summaries, and $\odot$ is the element-wise (Hadamard) product.
By including $\sigma(\bm{\theta})$, this ensures we can treat each of the components of $\bm{\Gamma}$ as the number of standard deviations that we are shifting the corresponding model summary statistic. This approach is similar to $\text{ABC}_{\mu}$, as it models the discrepancy at the level of the summary statistics. The adjustment parameters in RBSL-M are assigned a Laplace (double exponential) prior distribution, $\pi(\bm{\Gamma}) \sim \text{Laplace}(0, \lambda)$, which encourages sparsity and robustness. 
The Laplace prior concentrates most of its mass near zero, reflecting the assumption that no adjustment is needed in the absence of misspecification, while its heavy tails allow for significant adjustments when necessary. The hyperparameter $\lambda$ controls the spread of the prior, selecting an appropriate value ensures that the tails are sufficiently heavy to detect and correct for model misspecification. As a default choice, setting $\lambda=0.5$ is a reasonable trade-off. For a more principled approach, one could ideally use additional validation data and tune $\lambda$ appropriately, but if additional data is not available, the modeller can conduct posterior predictive checks to determine if there are large discrepancies between simulated and true summaries that are uncorrected.

In the RBSL-V variant, instead of adjusting the means, the variances are inflated to account for misspecification:
\begin{equation*}\label{eq:rbslv_var}
V(\bm{\theta}, \bm{\Gamma}) = \Sigma(\bm{\theta}) + \text{diag}\big( [\Sigma(\bm{\theta})]_{ii} \gamma_i^2 \big),
\end{equation*}
where $[\Sigma(\bm{\theta})]_{ii}$ is the ($i,i$)-th element of the covariance matrix $\Sigma(\bm{\theta})$.
 The adjusted covariance matrix $V(\bm{\theta}, \bm{\Gamma})$ is then used in the synthetic likelihood, $\mathcal{N}(S(\bm{y}) \mid \mu(\bm{\theta}), V(\bm{\theta}, \bm{\Gamma}))$.
The adjustment parameters $\gamma_i$ in RBSL-V are assigned an exponential prior, $\pi(\gamma_i) \sim \text{Exponential}(\lambda)$, allowing only positive values.
The augmented BSL posterior can be sampled using a component-wise MCMC algorithm, where the model parameters $\bm{\theta}$ are sampled via Metropolis-Hastings \citep{metropolis_equation_1953, hastings_monte_1970} and the adjustment parameters $\bm{\Gamma}$ are sampled via slice sampling \citep{neal_slice_2003}. The RBSL-M algorithm is presented in Algorithm~\ref{alg:rbsl}.
While we use the same adjustment parameter notation as RBSL-M for consistency, we can also interpret RBSL-V as combining the standard synthetic likelihood with an independent Gaussian error model which has a diagonal covariance matrix, where we treat the covariance terms as unknown parameters.

\begin{algorithm} \caption{ MCMC RBSL-M}\label{alg:rbsl} 
\begin{algorithmic}[1] 
\REQUIRE{Observed summary statistics $S(\bm{y})$; prior distribution $\pi(\bm{\theta})$; proposal kernel $\mbox{$q(\cdot \mid \cdot)$}$; number of simulations $n$; number of MCMC iterations $T$; initial values $\bm{\theta}^0$, $\bm{\Gamma}^0$
}
\ENSURE{MCMC samples $\{(\bm{\theta}^t, \bm{\Gamma}^t)\}_{t=1}^T$ from the RBSL-M posterior}
\STATE{Compute estimates $\widehat{\bm{\mu}}(\bm{\theta}^0)$ and $\widehat{\bm{\Sigma}}(\bm{\theta}^0)$ using $n$ simulated summary statistics} 
    \STATE{Compute $\widehat{g}(S(\bm{y}) \mid \bm{\theta}^0) = \mathcal{N}\bigl(S(\bm{y}) \mid \widehat{\bm{\mu}}(\bm{\theta}^0) + \widehat{\sigma}(\bm{\theta}^0) \odot \bm{\Gamma}^0,\ \widehat{\bm{\Sigma}}(\bm{\theta}^0) \bigr)$}
\FOR{$t=1$ to $T$} 
    \FOR{$j=1$ to $d$}
        \STATE{Update $\gamma_j^t$ with slice sampling}
    \ENDFOR
    \STATE{Propose $\bm{\theta}^* \sim q(\cdot \mid \bm{\theta}^{t-1})$}
    \STATE{Compute estimates $\widehat{\bm{\mu}}(\bm{\theta}^*)$ and $\widehat{\bm{\Sigma}}(\bm{\theta}^*)$ using $n$ simulated summary statistics}
        \STATE{Compute $\widehat{g}(S(\bm{y}) \mid \bm{\theta}^*) = \mathcal{N}\bigl(S(\bm{y}) \mid \widehat{\bm{\mu}}(\bm{\theta}^*) + \widehat{\sigma}(\bm{\theta}^*) \odot \bm{\Gamma}^t, \widehat{\bm{\Sigma}}(\bm{\theta}^*) \bigr)$}
    \STATE{Perform a Metropolis–Hastings step for $\bm{\theta}^*$ using $\pi(\bm{\theta}\mid\bm{\Gamma}^t,S(\bm{y}))$: accept $\bm{\theta}^*$ with probability $\alpha = \min{1, \frac{\pi(\bm{\theta}^*,\bm{\Gamma}^t \mid S(\bm{y})) q(\bm{\theta}^{t-1}\mid\bm{\theta}^*)}{\pi(\bm{\theta}^{t-1},\bm{\Gamma}^t\mid S(\bm{y})) q(\bm{\theta}^*\mid\bm{\theta}^{t-1})}}$}
    \IF{accepted}
        \STATE{$\bm{\theta}^t \leftarrow \bm{\theta}^*$}
    \ELSE
        \STATE{$\bm{\theta}^t \leftarrow \bm{\theta}^{t-1}$} 
    \ENDIF 
\ENDFOR 
\RETURN $\{(\bm{\theta}^t, \bm{\Gamma}^t)\}_{t=1}^T$
\end{algorithmic} 
\end{algorithm}


A lesson from RBSL-V is that modifying the covariance matrix in the synthetic likelihood can lead to more robust inference than the default covariance matrix used in the standard synthetic likelihood. 
Building on this, \citet{frazier_synthetic_2024}, propose using a fixed estimated covariance matrix $\Delta$ instead of the usual parameter-dependent covariance matrix $\Sigma(\bm{\theta})$ that is estimated.
The resulting posterior samples can then be adjusted, similar to RBSL-V, to ensure valid frequentist coverage, where coverage is based on the pseudo-true parameter as defined in ABC. 
Recently, \citet{nguyen_wasserstein_2023} combine RBSL with a Wasserstein Gaussianisation transformation and variational Bayes scheme. This allows for the use of non-Gaussian summary statistics and enables efficient scaling to higher-dimensional summaries through the computational efficiency of variational Bayes methods. 

Adjustment parameters have also been incorporated into neural methods. \citet{ward_robust_2022} propose robust neural posterior estimation (RNPE), which extends NPE to account for misspecification. After training an NPE posterior, a separate normalising flow, $q_{\varphi}(\bm{x})$, is trained over simulations from the prior predictive distribution.
To add robustness, simulations are adjusted using, $\tilde{\bm{x}}_m \sim \widehat{p}(\bm{x} \mid \bm{y}) \propto p(\bm{y} \mid \bm{x})q(\bm{x})$. \citet{ward_robust_2022} employ a spike-and-slab prior over the discrepancies, with the spike being a low-variance normal distribution—representing well-specified models, and the slab a high-variance Cauchy distribution—allowing for larger discrepancies. By introducing a Bernoulli variable for each summary statistic to indicate spike or slab membership, the posterior of this variable serves as an indicator of the misspecification probability for that summary statistic.
Posterior samples are obtained by passing the adjusted data into the trained NPE posterior, $\tilde{\bm{\theta}} \sim q(\bm{\theta} \mid \tilde{\bm{x}}_m)$. 
Similarly, \citet{kelly_misspecification-robust_2024} introduce robust sequential neural likelihood (RSNL), targeting an approximate joint posterior:
\begin{equation*}
\pi(\bm{\theta}, \bm{\Gamma} \mid S(\bm{y})) \propto q_{\bm{\phi^*}}(S(\bm{y}) - \bm{\Gamma} \mid \bm{\theta}) \pi(\bm{\theta}) \pi(\bm{\Gamma}),
\end{equation*}
where $q_{\bm{\phi^*}}$ is the learnt neural likelihood approximation and $\pi(\bm{\Gamma})$ is a Laplace prior adaptively set in each sequential round. This method allows adjustment parameters to be incorporated into neural likelihood approaches with the possibility of sequential sampling.

\FloatBarrier

\section{Illustrative example revisited}\label{sec:toy_robust}
To demonstrate how the robust methods discussed in Section~\ref{sec:methods} can improve inference under model misspecification, we revisit the misspecified MA(1) example from Section~\ref{sec:ill_example}. In the standard setting, both BSL and SNL gave poor inference, with the approximate posteriors concentrating in a region of the parameter space far from the pseudo-true parameter value. 

\begin{figure}[!htbp]
\centering
\begin{subfigure}[b]{0.49\textwidth}
    \includegraphics[width=\textwidth]{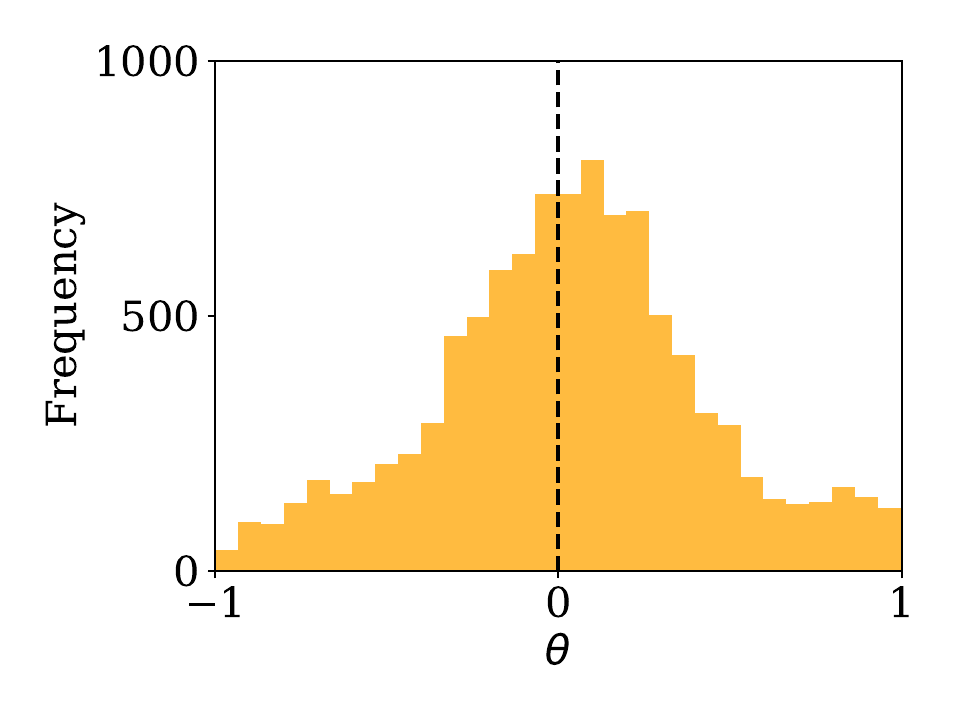}
    \caption{RBSL-M posterior}
    \label{fig:method3a}
\end{subfigure}
\hfill
\begin{subfigure}[b]{0.49\textwidth}
    \includegraphics[width=\textwidth]{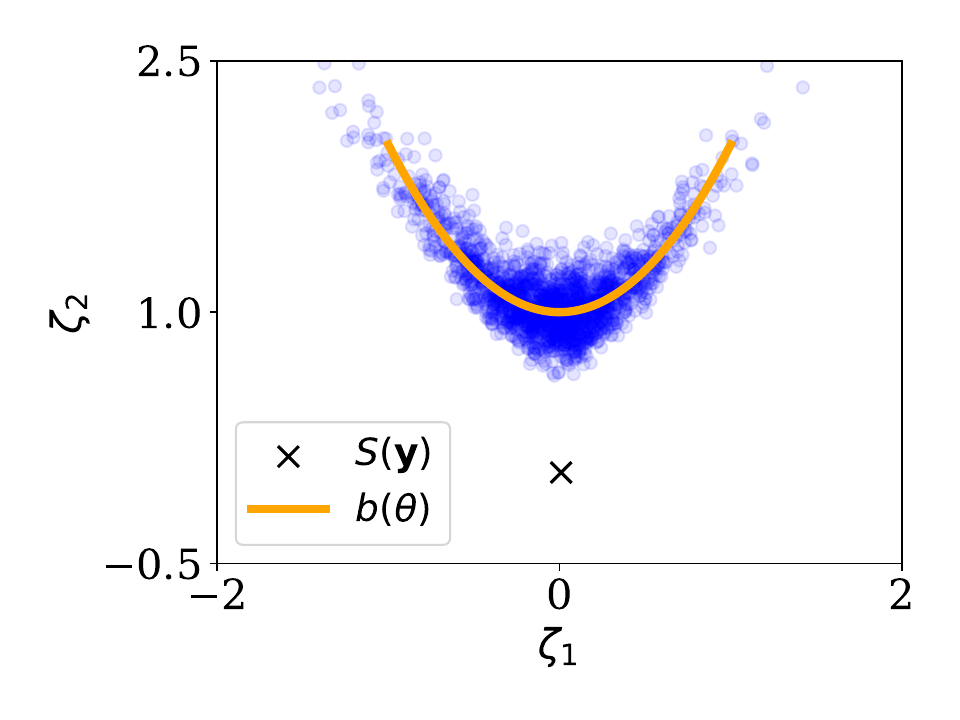}
    \caption{Posterior predictive for RBSL-M}
    \label{fig:method3b}
\end{subfigure} \\
\begin{subfigure}[b]{0.49\textwidth}
    \includegraphics[width=\textwidth]{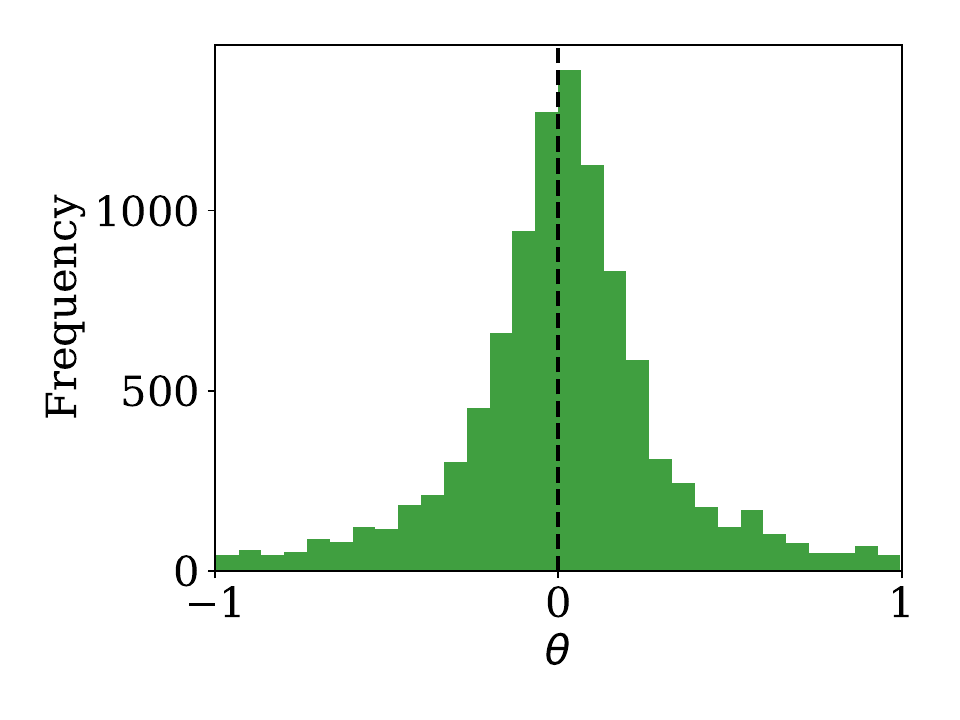}
    \caption{RBSL-V Posterior}
    \label{fig:method3a}
\end{subfigure}
\hfill
\begin{subfigure}[b]{0.49\textwidth}
    \includegraphics[width=\textwidth]{figs/rbslm_ppc.pdf}
    \caption{Posterior predictive for RBSL-V}
    \label{fig:method3b}
\end{subfigure}
\caption{Comparison of RBSL-M (top row) and RBSL-V (bottom row) results on the misspecified MA(1) example. Left panels: posterior distributions for $\theta$ with the pseudo-true value at $\theta=0$ indicated by a vertical dashed line. Right panels: posterior predictive summaries (circles) compared to the observed summary ($\times$) and the binding function $b(\theta)$ (solid curve).}
\label{fig:misspec_ma1_rbsl}
\end{figure}

A straightforward way to address misspecification here is to remove the problematic summary, guided by either prior or posterior predictive checks. In this toy example, the Bayesian workflow should detect the issue immediately in the prior predictive step (see the left panel of Figure~\ref{fig:misspec_ma1_abc}). However, if the prior is very broad and resources are limited, one might run the chosen SBI algorithm first, then use posterior predictive checks. In this example, these posterior predictive checks also show that $\zeta_1$ can be matched, but $\zeta_2$ cannot (e.g., evident in posterior predictive plots in Figure~\ref{fig:misspec_ma1_rbsl}). 
We nevertheless keep the problematic summary henceforth for pedagogical reasons. Such visual inspections may not suffice for more complex scenarios—particularly if some summaries are matched marginally but not jointly—but the diagnostics discussed in Section~\ref{sec:sbi_mm} can help detect these cases, and robust methods can then be applied.

\begin{figure}[htbp!]
    \centering
    \includegraphics[width=0.5\linewidth]{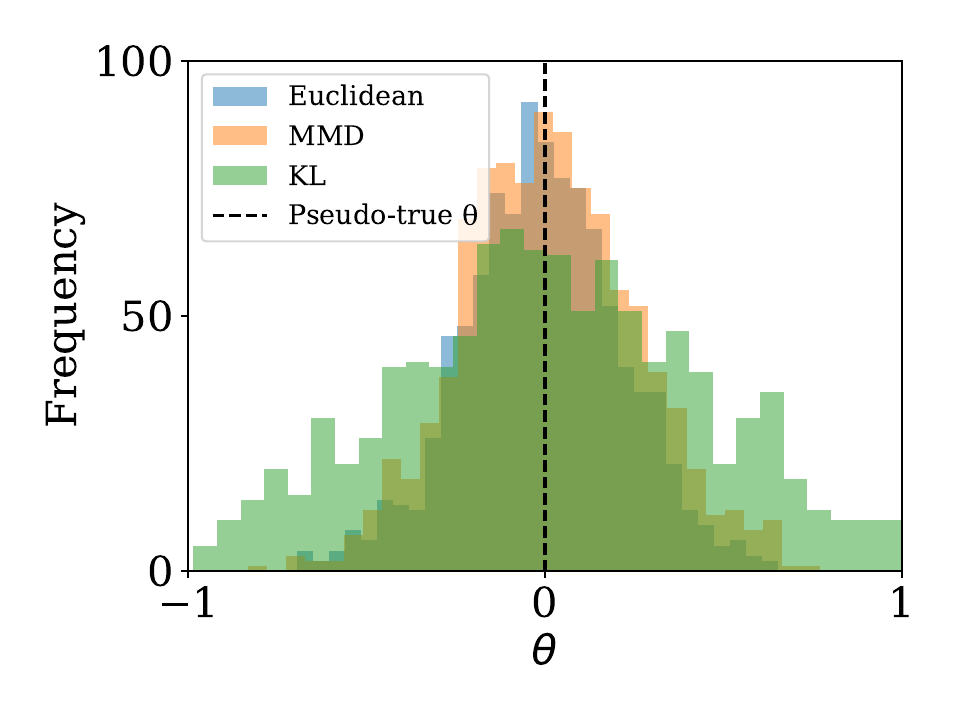}
    \vspace{-3mm}
    \caption{Approximate ABC posteriors under three discrepancy choices—Euclidean distance, KL divergence, and MMD. The dashed line marks the pseudo-true value at $\theta = 0$.}
    \label{fig:abc_losses}
\end{figure}

We illustrate approximate posteriors in rejection ABC under three commonly used discrepancy choices—Euclidean distance, KL divergence, and MMD—in Figure~\ref{fig:abc_losses}.
These results use the full dataset without summarisation. Notably, the posterior with KL divergence is broader, while those with MMD and Euclidean distance are more tightly centred around the pseudo-true value at $\theta=0$. These findings align with \citet{legramanti_concentration_2025} and reinforce the view of MMD being a robust choice of loss when we view ABC as a form of generalised Bayesian inference.

\begin{figure}[!htbp]
\centering
\begin{subfigure}[b]{0.49\textwidth}
    \includegraphics[width=\textwidth]{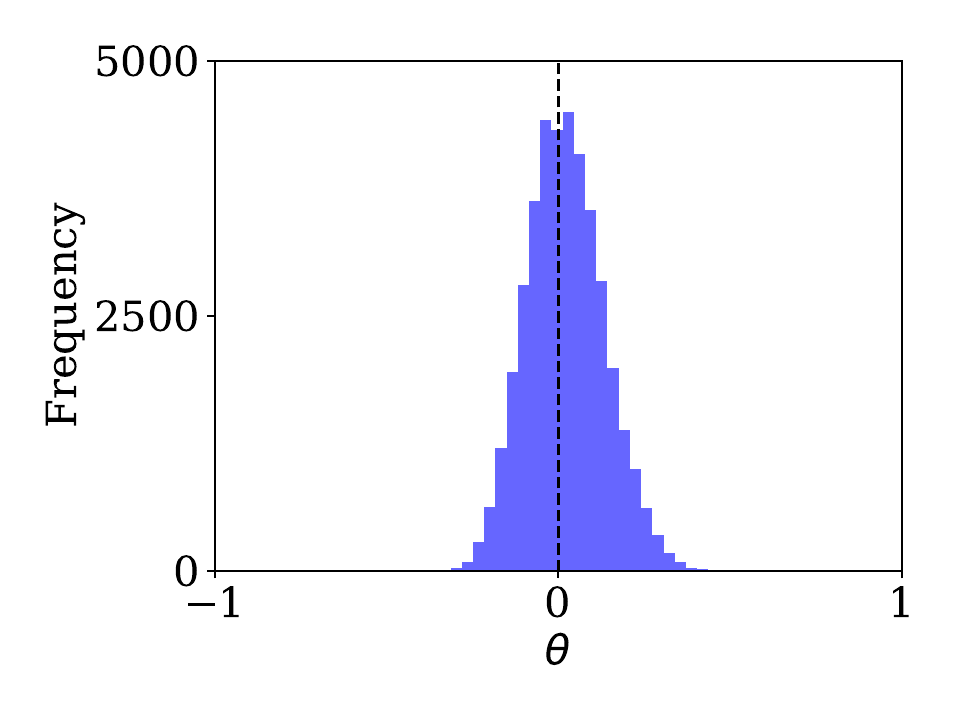}
    \caption{RSNL posterior}
    \label{fig:method2a}
\end{subfigure} 
\hfill
\begin{subfigure}[b]{0.49\textwidth}
    \includegraphics[width=\textwidth]{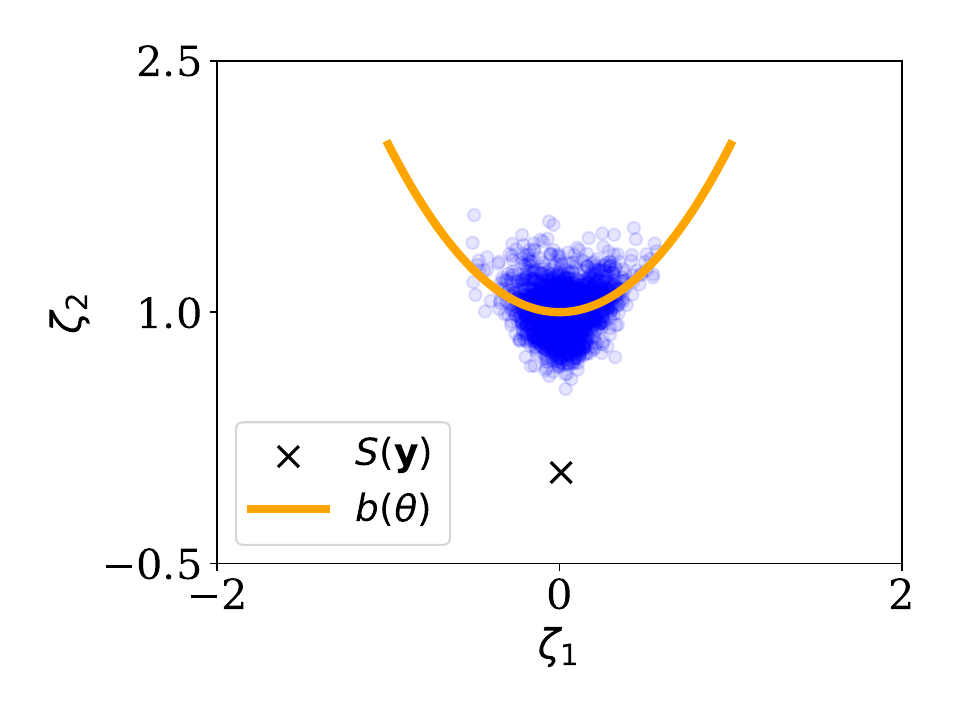}
    \caption{Posterior predictive for RSNL.}
    \label{fig:method2b}
\end{subfigure}
\caption{RSNL results on the misspecified MA(1) example. Left panel: Posterior distribution for $\theta$, with the pseudo-true value at $\theta = 0$ indicated by a vertical dashed line. Right panel: Posterior predictive summaries (circles) compared to the observed summary ($\times$) and the binding function $b(\theta)$ (solid curve).}
\label{fig:misspec_ma1_rsnl}
\end{figure}

By incorporating adjustment parameters, we can achieve more robust inference. Figure~\ref{fig:misspec_ma1_rbsl} illustrates the results for RBSL-M and RBSL-V. Compared to standard BSL, these robust variants produce posteriors centred more closely around the pseudo-true parameter value at $\theta = 0$. Their posterior predictive simulations also better match the observed summaries, mitigating the previously observed overconfidence and poor coverage. We see similar improvements for neural methods when adopting robust strategies, as shown in Figure~\ref{fig:misspec_ma1_rsnl} for RSNL.

\begin{figure}[!htbp]
\centering
\begin{subfigure}[b]{0.49\textwidth}
    \includegraphics[width=\textwidth]{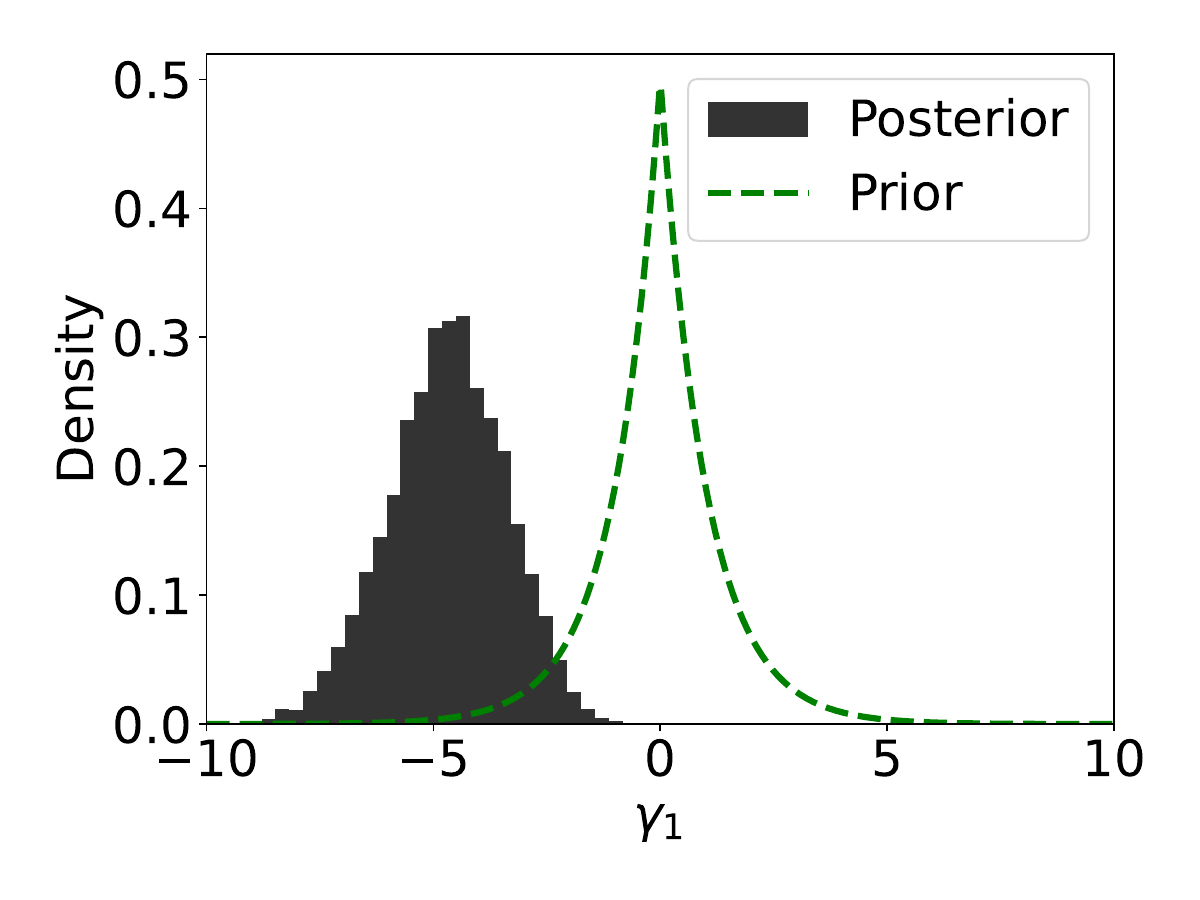}
    \label{fig:method3a}
\end{subfigure}
\hfill
\begin{subfigure}[b]{0.49\textwidth}
    \includegraphics[width=\textwidth]{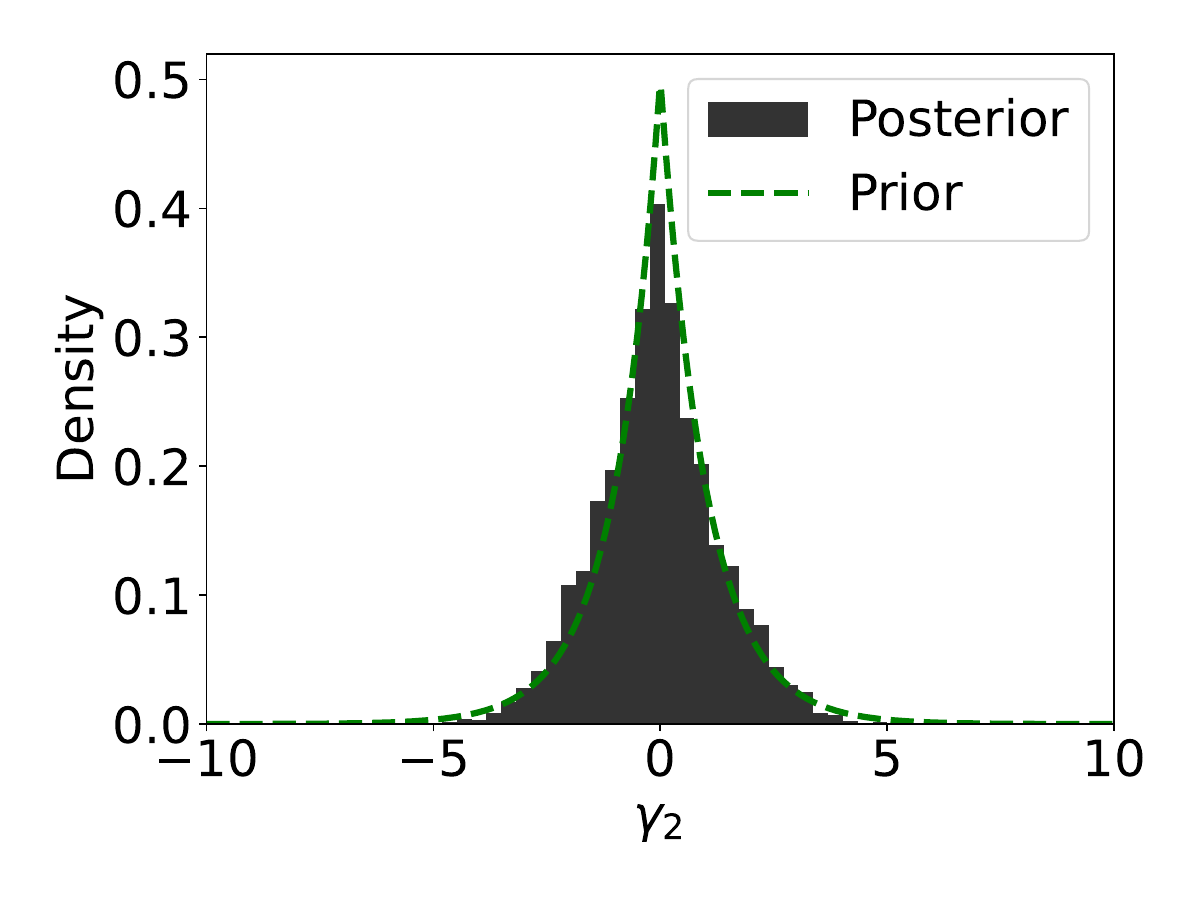}
    \label{fig:method3b}
\end{subfigure} \\
\begin{subfigure}[b]{0.49\textwidth}
    \includegraphics[width=\textwidth]{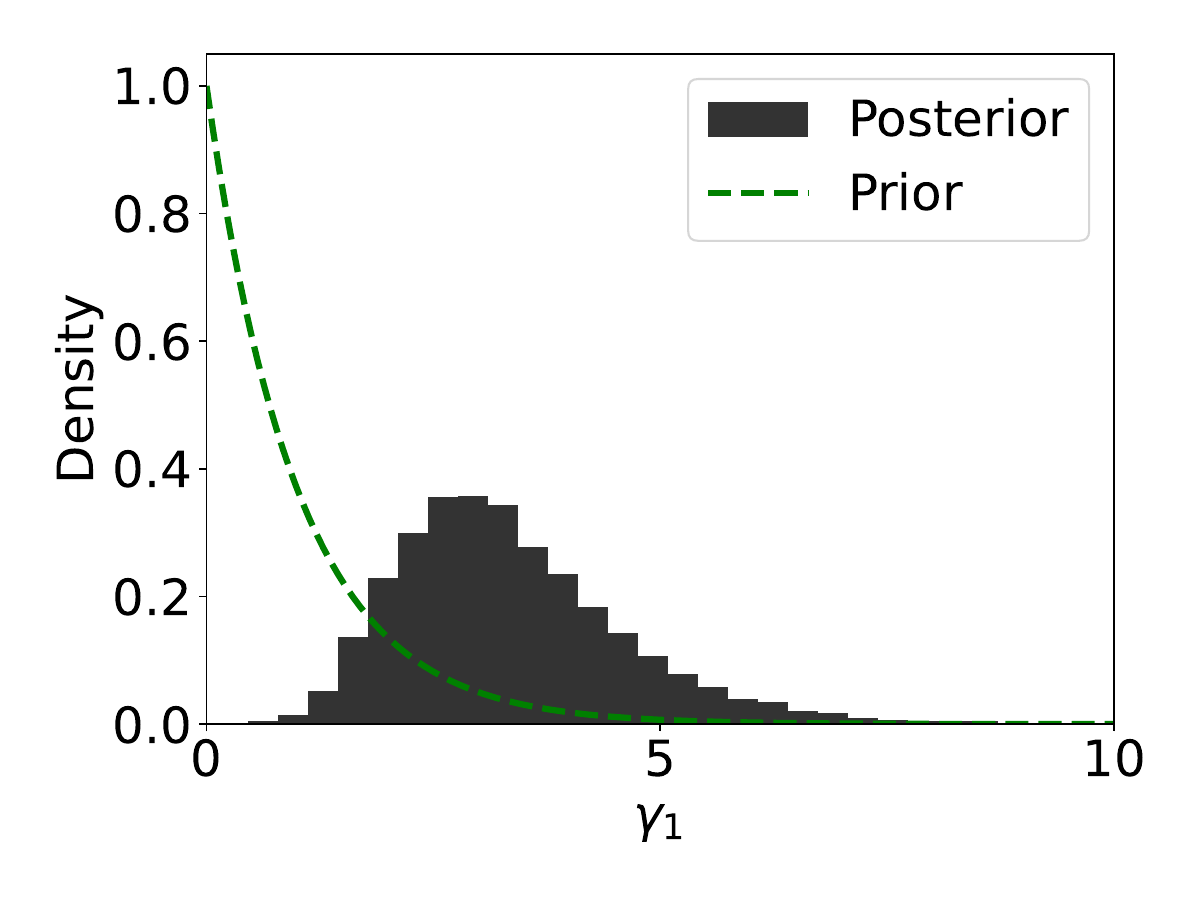}
    \label{fig:method3a}
\end{subfigure}
\hfill
\begin{subfigure}[b]{0.49\textwidth}
    \includegraphics[width=\textwidth]{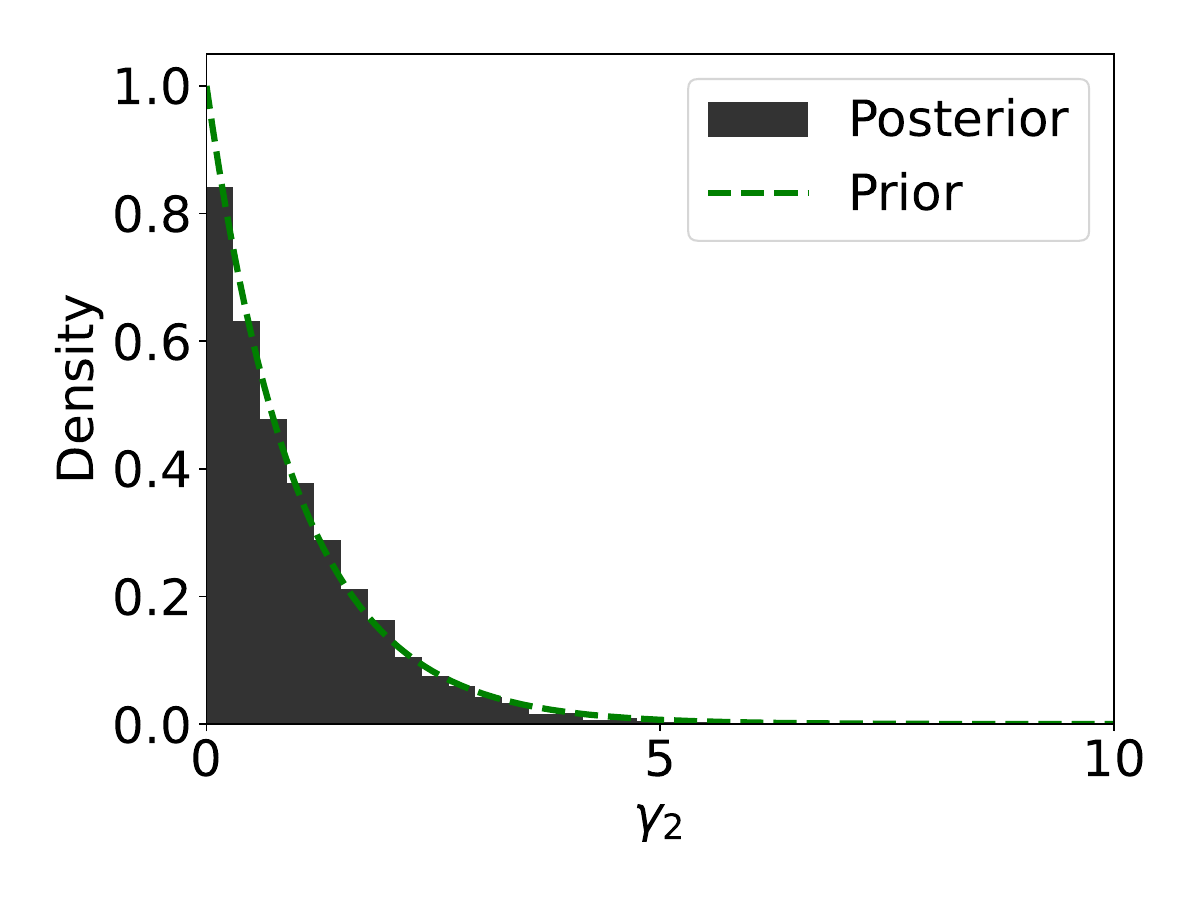}
    \label{fig:method3b}
\end{subfigure}

\caption{Posterior distributions of the adjustment parameters under RBSL-M (top row) and RBSL-V (bottom row) for the misspecified MA(1) example. Each panel corresponds to one component of $\bm{\Gamma}$. The dashed lines indicate the prior density, while the histograms summarise samples from the posterior distribution of the adjustment parameters.}
\label{fig:adjustment_param_plots}
\end{figure}

An additional advantage of using adjustment parameters is for model criticism. Figure~\ref{fig:adjustment_param_plots} shows that the posterior for the first component of $\bm{\Gamma}$, $\gamma_1$, differs noticeably from its prior distribution, flagging that the model is incompatible with the first summary statistic. This suggests that the chosen MA(1) model cannot replicate the observed sample variance. While this is a simple example, the same logic applies to more complex scenarios, helping modellers pinpoint and address the aspects of their models that fail to capture key features of the data. 

\FloatBarrier






\section{Discussion}\label{sec:discussion}

Model misspecification has now been established as a central issue in SBI. In this paper, we have provided a comprehensive overview of model misspecification in SBI, covering how it affects key methods—approximate Bayesian computation (ABC), Bayesian synthetic likelihood (BSL), and neural conditional density estimation (NCDE)—and outlined recent methods that address model misspecification. 

Constructing robust models in practice requires practitioners to be aware of model misspecification and to iteratively refine their models. Within a principled Bayesian workflow, model checking plays a crucial role. Diagnostics such as posterior predictive checks, or more specific SBI-focused diagnostics, such as those covered at the end of Section~\ref{sec:mm_sbi}, can help reveal where and how a model fails to capture key aspects of the data. Informed by these diagnostics, one can iteratively improve the model or incorporate robust inference techniques.

In Section~\ref{sec:methods}, we outlined three broad strategies for robust inference in SBI: robust summary statistics, generalised Bayesian inference (GBI), and adjustment parameters. Employing robust summary statistics—such as medians rather than means—can reduce sensitivity to outliers and minor discrepancies. While many automated approaches to learning summaries are susceptible to misspecification, there is active work in robust automated summary construction, such as the approach proposed by \citealp{huang_learning_2023}. For SBI methods relying on distance measures between simulated and observed data, opting for robust metrics (e.g., MMD-based approaches) can confer resilience. Testing multiple distance metrics, where feasible, may provide further insights. Similarly, if using BSL, NPE, or SNL, employing robust adjustment parameters can enhance reliability with minimal computational overhead.
By categorising recent developments, we might consider whether multiple of these robust strategies could be employed together, integrated into a Bayesian workflow, to further strengthen robustness and improve the overall quality of inference.

While we do not explicitly compare the robustness properties of ABC, BSL, and NCDE here, it is worth noting that standard ABC methods often exhibit a degree of inherent robustness \citep{schmon_generalized_2020}. Ironically, features of ABC sometimes viewed as drawbacks—such as reliance on summary statistics, error tolerance thresholds, and a user-specified discrepancy function—can, in fact, mitigate certain forms of misspecification. For instance, robust summaries (Section~\ref{sec:robust_summaries}) may be less sensitive to outliers than using the full dataset, tolerance thresholds effectively act as implicit error models \citep{miller_robust_2019, wilkinson_approximate_2013}, and choosing a robust distance (Section~\ref{sec:gbi}) can reduce the influence of outliers compared to standard Bayesian inference. Although ABC scales poorly to high-dimensional data, its natural resilience may make it preferable for low-dimensional problems.

Despite recent advances, many open questions remain and point to exciting avenues for future work. One priority is the development of standardised metrics and benchmarks for misspecified models, similar to the benchmarks of \citet{lueckmann_benchmarking_2021} in the well-specified setting. Such benchmarks would facilitate the comparison of different methods and provide guidelines for practitioners. Another important gap lies in the theoretical foundations for NCDE under misspecification. While ABC and BSL have benefited from rigorous theory of their behaviour under misspecification, NCDE approaches currently lack a similar level of theoretical clarity.
 Recent theoretical insights into NCDE methods rely on the compatibility assumption \citep{frazier_statistical_2024}, but the case for model misspecification is less clear and warrants further investigation.

For sequential sampling in SBI, parameter samples are intended to be drawn in regions of higher density, but standard approaches can be unstable. One challenge is ``leakage'' of posterior mass outside the prior support \citep{durkan_contrastive_2020}, for which truncated prior proposals have been suggested \citep{deistler_truncated_2022}. Another issue arises when extreme prior predictive samples degrade training, which can be mitigated by preconditioned NPE, which uses an initial training dataset of ABC samples \citep{wang_preconditioned_2024}.
Under model misspecification, the poor empirical performance of neural SBI approaches suggests that they do not naturally converge to a suitable pseudo-true parameter, but it has been well-established that ABC can. Hence, sequential approaches that have been preconditioned with ABC samples may have improved robustness in misspecified settings.

The challenge of obtaining posteriors with credible intervals that align with true coverage probabilities is a well-established challenge in SBI, with many NCDE methods prone to giving overconfident inference \citep{hermans_crisis_2022}. Such issues are likely to be exacerbated when the model is not correctly specified, as evident in the empirical results in \citet{cannon_investigating_2022}, and by the general observation that Bayesian credible sets are not valid confidence sets when the model is misspecified \citep{kleijn_bernstein-von-mises_2012}. Addressing calibration under misspecification may benefit from recent work that leverages optimal transport theory to achieve more robust and calibrated inference \citep{wehenkel_addressing_2024}. Meanwhile, emerging classes of SBI methods, such as flow-matching and diffusion models \citep{gloeckler_all--one_2024, simons_neural_2023, wildberger_flow_2023}, remain largely unexplored in the context of misspecification. Investigating their robustness and designing methods to mitigate model misspecification will likely expand their practical applicability. Further, training neural approximations that, instead of the usual loss minimising forward KL divergence, consider the generalised variational inference (GVI) framework \citep{knoblauch_optimization-centric_2022} may also help learn more robust approximations.

The field of SBI is beginning to seriously engage with the problem of model misspecification. Many robust methods have now been developed, and the theoretical understanding of their behaviour under model misspecification is steadily improving. Neural methods, though popular for their scalability, are sensitive to misspecification. Lessons learned from the inherent robustness of ABC might inspire further robust versions of NCDE approaches. Ultimately, practitioners must be vigilant about model misspecification, conducting model checking, and incorporating robust methods when needed. We anticipate that ongoing research will continue to expand the suite of available tools, making misspecification-robust inference increasingly achievable.

\section*{Acknowledgements}
RPK was supported by a PhD Research Training Program scholarship from the Australian Government and a QUT Centre for Data Science top-up scholarship. 
CD and DTF were supported by Australian Research Council funding schemes FT210100260 and DE200101070, respectively.
DJN was supported by the Ministry of Education, Singapore, under the Academic Research Fund Tier 2 (MOE-T2EP20123-0009) grant.
RPK, DJW, and CD thank the Centre for Data Science at QUT for its support. 

\bibliographystyle{apalike}
\bibliography{references}       


\end{document}